\documentclass[%
 reprint,
 superscriptaddress,
 amsmath,%amssymb,
 aps,
 prb,
 longbibliography
]{revtex4-2}

\usepackage{graphicx, color}% Include figure files
\usepackage{newtxtext, newtxmath}
\usepackage{bm}% bold math
\usepackage{braket, mathtools}
\usepackage{diagbox}
\usepackage[
 colorlinks=true,
 citecolor=blue,
 linkcolor=blue
]{hyperref}

\graphicspath{{./images/}}

\DeclareMathOperator{\tr}{tr}
\DeclareMathOperator{\diag}{diag}

\newcommand{\ii}{\mathrm{i}}
\newcommand{\ee}{\mathrm{e}}

\begin{document}

\title{%
 \texorpdfstring%
 {Fulde--Ferrell--Larkin--Ovchinnikov state induced by antiferromagnetic order \\ in $\kappa$-type organic conductors}%
 {Fulde--Ferrell--Larkin--Ovchinnikov state induced by antiferromagnetic order in κ-type organic conductors}
}
%\thanks{A footnote to the article title}

\author{Shuntaro Sumita}
\email[]{s-sumita@g.ecc.u-tokyo.ac.jp}
\affiliation{%
 Department of Basic Science, The University of Tokyo, Meguro, Tokyo 153-8902, Japan
}%
\affiliation{%
 Komaba Institute for Science, The University of Tokyo, Meguro, Tokyo 153-8902, Japan
}%
\affiliation{%
 Condensed Matter Theory Laboratory, RIKEN CPR, Wako, Saitama 351-0198, Japan
}%

\author{Makoto Naka}
% \email[]{m-naka@mail.dendai.ac.jp}
\affiliation{%
 School of Science and Engineering, Tokyo Denki University, Ishizaka, Saitama 350-0394, Japan
}%

\author{Hitoshi Seo}
% \email[]{seo@riken.jp}
\affiliation{%
 Condensed Matter Theory Laboratory, RIKEN CPR, Wako, Saitama 351-0198, Japan
}%
\affiliation{%
 RIKEN Center for Emergent Matter Science, Wako, Saitama 351-0198, Japan
}%

\date{\today}

\begin{abstract}
 We theoretically investigate superconductivity under a spin-split band structure owing to a collinear-type antiferromagnetic order in quasi-two-dimensional organic compounds $\kappa$-(BEDT-TTF)$_2X$.
 We find that the magnetic order can induce a Fulde--Ferrell--Larkin--Ovchinnikov (FFLO) state, where the Cooper pair possesses a finite center-of-mass momentum.
 We show this from two types of analyses: (1) an effective model where simple intraband attractive interactions are assumed, and (2) many-body calculations of the repulsive Hubbard model based on the fluctuation-exchange approximation and the linearized Eliashberg equation.
 Our results show the possibility of realizing the FFLO state without applying an external magnetic field.
\end{abstract}

\maketitle

\section{Introduction}
The relationship between magnetism and superconductivity has been of major interest in the field of strongly correlated electron systems.
In particular, a coexistent state of the two orders can symptomize unconventional superconductivity since conventional BCS-type superconductors do not favor magnetism, resulting in an exotic state.
For example, UGe$_2$~\cite{Saxena2000}, UCoGe~\cite{Huy2007}, and URhGe~\cite{Aoki2001} are known as representative spin-triplet superconductors coexisting with a ferromagnetic (FM) order~\cite{[][{, and references therein.}]{Aoki2019_review}}.
Another example is symmetry-protected nodal superconductivity in antiferromagnetic (AFM) systems~\cite{[][{, and references therein.}]{Sumita2022_review}}, e.g., UPd$_2$Al$_3$~\cite{Nomoto2017} and Sr$_2$IrO$_4$~\cite{Sumita2017}.

\begin{figure}
 \includegraphics[width=.8\linewidth, pagebox=artbox]{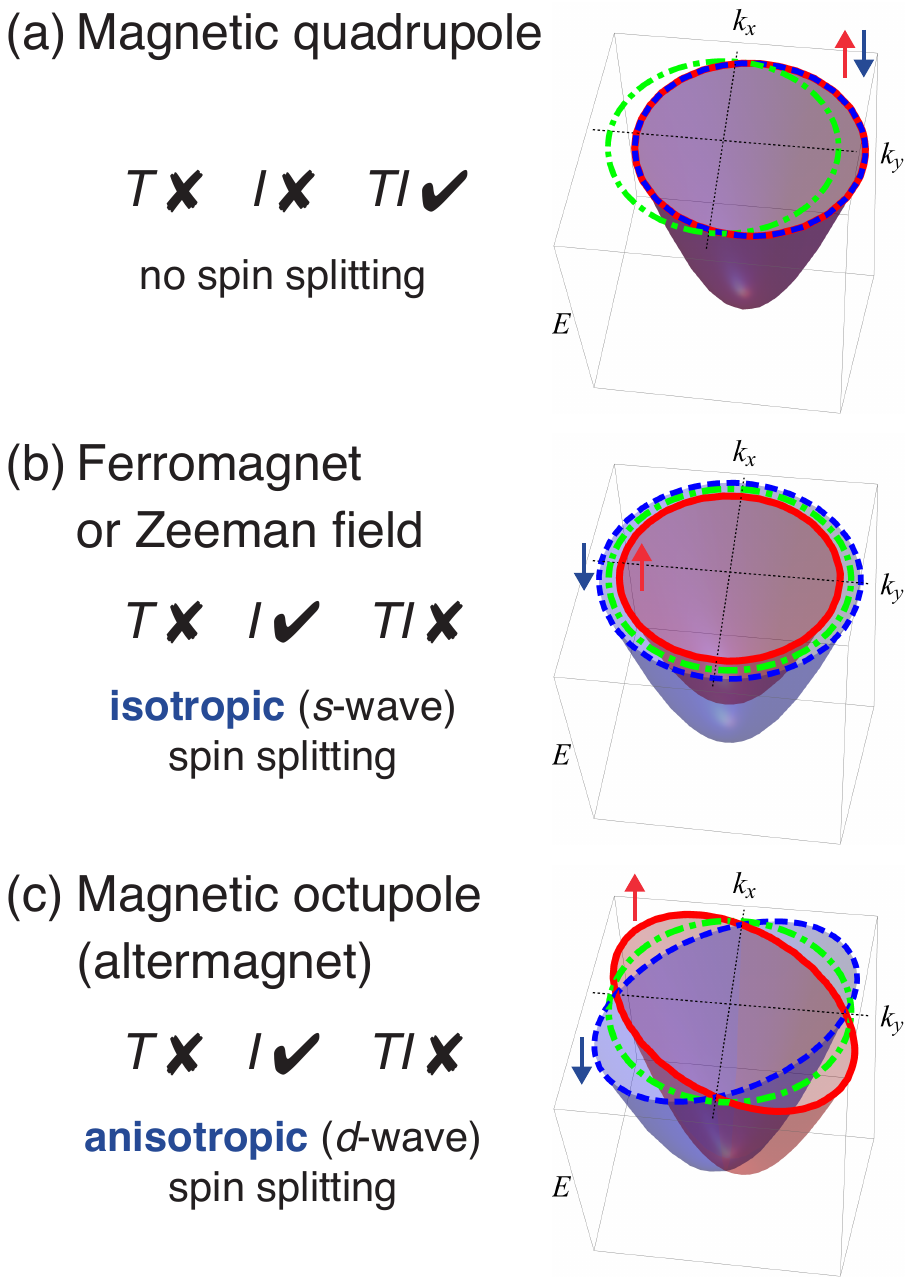}
 \caption{Symmetry properties and schematic energy band structures in (a) magnetic quadrupole, (b) FM (or Zeeman field), and (c) magnetic octupole states. The red and blue dispersions represent (pseudo)spin up and down bands, respectively. The green dash-dotted lines are the Fermi surface in the paramagnetic state.}
 \label{fig:magnetism_and_band}
\end{figure}

Interestingly, theoretical studies showed that Fulde--Ferrell--Larkin--Ovchinnikov (FFLO) superconductivity~\cite{FF, LO} with finite center-of-mass (COM) momentum of the Cooper pairs can be realized under an odd-parity magnetic multipole order~\cite{Sumita2016, Sumita2017, Wu2023_arXiv, Amin2023_arXiv}, such as in the AFM state in locally noncentrosymmetric systems~\cite{Yanase2014, HikaruWatanabe2017}.
In this case, the odd-parity magnetic quadrupole order breaks both spatial inversion ($I$) and time-reversal ($T$) symmetries, but preserves the combination of them.
As a result, the energy band structure is asymmetric for the momentum flip $\bm{k} \to -\bm{k}$, while up and down spins or pseudospins are degenerate at any $\bm{k}$ point [Fig.~\ref{fig:magnetism_and_band}(a)].
Note that the symmetry property is different from that in typical AFM systems, where magnetic translation symmetry (the combination of time-reversal and half-transition) is conserved.

Here we focus on another type of AFM order that breaks $T$ symmetry, which has been theoretically pointed out in $\kappa$-(ET)$_2$Cu[N(CN)$_2$]Cl (ET is the abbreviation of BEDT-TTF $=$ bis(ethylenedithio)tetrathia-fulvalene), abbreviated as $\kappa$-Cl in the following~\cite{Naka2019, Naka2020, Hayami2020}.
$\kappa$-Cl is a member of quasi-two-dimensional organic charge transfer salts $\kappa$-(ET)$_2X$ ($X$: monovalent anions), which is known as a platform of Mott physics; they typically show a collinear-type AFM insulating state or superconductivity at the ground state, which depends on the choice of the anion $X$, working as chemical pressure, and the applied physical pressure~\cite{McKenzie1997, Kanoda1997, Miyagawa2002}.
Because of their particular molecular arrangement [Fig.~\ref{fig:lattice_electronic_structure}(a)], the AFM order breaks $T$ symmetry but preserves $I$ symmetry, which induces spin splitting in the band structure.
The momentum-flip symmetry properties in the AFM state are the same as those in a FM state (or in a Zeeman field), whereas the band structure has \textit{anisotropic} spin splitting in contrast to \textit{isotropic} one in the FM state [see Figs.~\ref{fig:magnetism_and_band}(b) and \ref{fig:magnetism_and_band}(c)].
The AFM structure in $\kappa$-(ET)$_2X$ is classified into a ferroic ($\bm{q} = 0$) order of an even-parity magnetic octupole~\cite{Suzuki2017, Suzuki2019, Hayami2018, HikaruWatanabe2018} or a so-called altermagnetic order~\cite{Smejkal2022_Sep, Smejkal2022_Dec}, which is discussed also in inorganic compounds, not only in collinear-type AFM materials such as RuO$_2$~\cite{Ahn2019, Smejkal2020} and MnTe~\cite{Mazin2023} but also in coplanar magnets such as Mn$_3$Sn~\cite{Nakatsuji2015}.
Here we use the term AFM, instead of magnetic octupole or altermagnetic, following the convention of the research on $\kappa$-(ET)$_2X$.
Although the interplay between superconductivity and magnetism accompanied by the anisotropic spin splitting has attracted much attention~\cite{Soto-Garrido2014, Lee2021, Mazin2022_arXiv, Ouassou2023, Zhang2023_arXiv, Sun2023, Papaj2023, Zhu2023, Ghorashi2023_arXiv, Beenakker2023, Wei2023_arXiv, Brekke2023_arXiv, Lee2023_arXiv, Giil2023_arXiv, Chakraborty2023_arXiv}, the microscopic understanding for the coexisting state is lacking.

In this paper, motivated by these backgrounds, we analyze a microscopic model for the $\kappa$-type salts under the unusual AFM order and seek for possible superconductivity.
We consider two cases based on the tight-binding model incorporating the AFM order, one introducing attractive interactions by hand and another more realistic model, and calculate conditions to stabilize superconducting (SC) state with anisotropic pairing symmetries discussed in the literature.
In both cases, we find that FFLO superconductivity with finite COM momentum is stabilized by the AFM order with spin splitting.
The FFLO state is originally proposed to be stabilized by the Zeeman field [Fig.~\ref{fig:magnetism_and_band}(b)], and has actually been reported in organic compounds such as $\kappa$-type~\cite{Singleton2000, Uji2006, Lortz2007, Bergk2011, Wright2011, Agosta2012, Mayaffre2014, Tsuchiya2015, Agosta2017}, $\lambda$-type~\cite{Tanatar2002, Coniglio2011, Uji2012, Uji2013, Uji2015, Imajo2021}, $\beta''$-type~\cite{Cho2009, Beyer2012, Koutroulakis2016, Uji2018, Sugiura2019_nQM, Sugiura2019_PRB} salts, and also in Sr$_2$RuO$_4$~\cite{Kikugawa2016, Kinjo2022}, iron-based superconductors~\cite{Terashima2013, Cho2017, Kasahara2020}, and heavy-fermion systems~\cite{Matsuda2007_review, Radovan2003, Bianchi2003, Kenzelmann2008, Kitagawa2018}.
On the other hand, the AFM-induced FFLO state is feasible in \textit{zero magnetic field and zero net magnetization}.

The paper is organized as follows.
In Sec.~\ref{sec:model} we explain the two-dimensional (2D) tight-binding Hamiltonian of $\kappa$-(ET)$_2X$ with an AFM molecular field, and show the symmetry and band structure of the model.
In Sec.~\ref{sec:effective_model} we analyze an effective model with intraband attractive interactions as a first step, which helps us intuitively understand the AFM-induced FFLO superconductivity.
Next, we discuss a more realistic situation, namely, a repulsive Hubbard model, and analyze it based on a fluctuation-exchange (FLEX) approximation and Eliashberg theory (Sec.~\ref{sec:Eliashberg_Hubbard}).
Finally, a discussion and a summary are given in Secs.~\ref{sec:discussion} and \ref{sec:summary}, respectively.

\section{Model and electronic band structure}
\label{sec:model}
\begin{figure*}
 \includegraphics[width=\linewidth, pagebox=artbox]{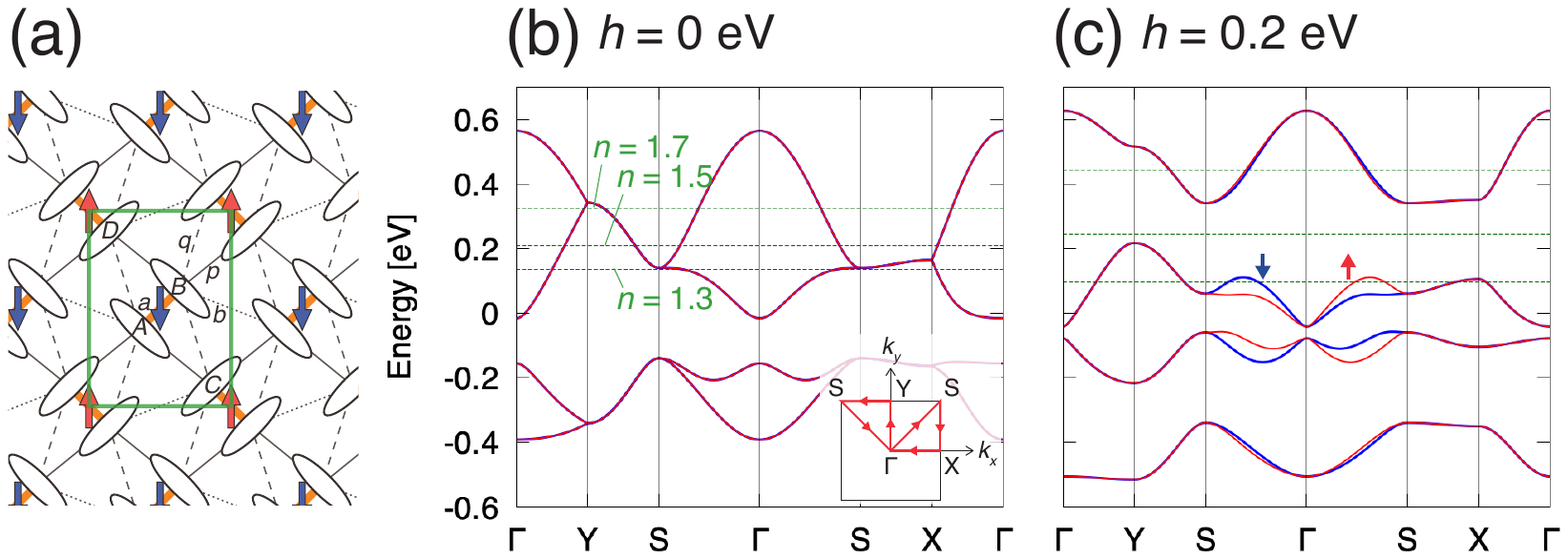}
 \caption{(a) Schematic view of the molecular arrangement in the conducting layer of $\kappa$-(ET)$_2X$. The unit cell (green rectangle) contains four ET molecules labeled by $A$--$D$. The bonds with large hoppings are shown: $a$ (orange bold), $b$ (dotted), $p$ (solid), and $q$ (dashed). The red and blue arrows represent up and down spin moments, respectively, in the AFM phase. The noninteracting band structures for (b) $h = 0$ and (c) $h = 0.2$ eV. Energy bands of up- and down-spin electrons are represented by the red and blue lines, respectively. The green dashed lines show the Fermi levels for the electron density $n = 1.3$, $1.5$, and $1.7$ per ET molecule. The inset in (b) represents the path in the 2D Brillouin zone. Note that, in the orthorhombic wallpaper group $p2gg$, $k_x = k_y = \pi$ is labeled by S, which corresponds to the M point in previous studies~\cite{Kino1996, Naka2019, Naka2020, Hayami2020}.}
 \label{fig:lattice_electronic_structure}
\end{figure*}

First, we introduce a 2D tight-binding model under the AFM ordering of $\bm{q} = 0$~\cite{Kino1996, Seo2000, Seo2004, Hayami2020}.
The conducting layer of $\kappa$-(ET)$_2X$ has four independent ET molecules $A$--$D$ in the unit cell, where $A$--$B$ and $C$--$D$ form dimers with different orientations [Fig.~\ref{fig:lattice_electronic_structure}(a)].
By considering the frontier orbitals of the four molecules, the noninteracting Hamiltonian with the four sublattices in the unit cell is written as
\begin{gather}
 H_0 = \frac{1}{N} \sum_{\bm{k}} \bm{C}_{\bm{k}}^\dagger \hat{H}_0(\bm{k}) \bm{C}_{\bm{k}},
 \label{eq:Hamiltonian} \\
 \bm{C}_{\bm{k}} = [c_{\bm{k}, A \uparrow}, c_{\bm{k}, A \downarrow}, c_{\bm{k}, B \uparrow}, c_{\bm{k}, B \downarrow}, c_{\bm{k}, C \uparrow}, c_{\bm{k}, C \downarrow}, c_{\bm{k}, D \uparrow}, c_{\bm{k}, D \downarrow}]^{\text{T}},
\end{gather}
where $c_{\bm{k}, l s}$ is the annihilation operator of electrons carrying momentum $\bm{k}$ and spin $s$ on the sublattice $l$.
The operators are defined by the Fourier transformation of the real-space operators: for example, the annihilation operators are given as $c_{\bm{k}, l s} = \sum_{\bm{R}} \ee^{-\ii\bm{k}\cdot\bm{R}} c_{\bm{R}, l s}$, where $\bm{R}$ represents the position of the unit cell.
In this convention, internal coordinates of the four ET molecules are neglected; we will address this point at the end of this section.
The momentum-dependent Hamiltonian matrix is composed of two terms,
\begin{equation}
 \hat{H}_0(\bm{k}) = \hat{H}_{\text{kin}}(\bm{k}) + \hat{H}_{\text{AFM}}.
 \label{eq:Hamiltonian_momentum}
\end{equation}
The first term is the kinetic energy with the intradimer hopping ($t_a$) and interdimer hoppings $(t_b, t_p, t_q)$ [see Fig.~\ref{fig:lattice_electronic_structure}(a)],
\begin{widetext}
 \begin{equation}
  \hat{H}_{\text{kin}}(\bm{k}) =
  \begin{bmatrix}
   0 & t_a + t_b \ee^{-\ii k_x} & t_p (1 + \ee^{-\ii k_x}) & t_q (1 + \ee^{-\ii k_y}) \\
   t_a + t_b \ee^{\ii k_x} & 0 & t_q (1 + \ee^{\ii k_y}) & t_p (1 + \ee^{\ii k_x}) \\
   t_p (1 + \ee^{\ii k_x}) & t_q (1 + \ee^{-\ii k_y}) & 0 & (t_a \ee^{\ii k_x} + t_b) \ee^{-\ii k_y} \\
   t_q (1 + \ee^{\ii k_y}) & t_p (1 + \ee^{-\ii k_x}) & (t_a \ee^{-\ii k_x} + t_b) \ee^{\ii k_y} & 0
  \end{bmatrix}
  \otimes \hat{\sigma}_0,
  \label{eq:Hamiltonian_momentum_kin}
 \end{equation}
\end{widetext}
which is the Kronecker product of the $4 \times 4$ matrix in the sublattice space and the $2 \times 2$ identity matrix $\hat{\sigma}_0$ in the spin space.
The lattice constants are set to unity.
The second term in Eq.~\eqref{eq:Hamiltonian_momentum} represents a molecular field of the AFM order that breaks $T$ symmetry,
\begin{equation}
 \hat{H}_{\text{AFM}} = h \hat{M}_{\text{AF}} \otimes \hat{\sigma}_z,
\end{equation}
where $\hat{M}_{\text{AF}} := \diag[1, 1, -1, -1]$ and $\hat{\sigma}_i$ ($i = x, y, z$) are the matrix representing the antiferroic order between the $A$--$B$ and $C$--$D$ sublattices and the Pauli matrix in the spin space, respectively.
Throughout this paper, we use the hopping parameters $(t_a, t_b, t_p, t_q) = (-0.207, -0.067, -0.102, 0.043)$ eV obtained from a first-principles calculation for $\kappa$-Cl at 15 K~\cite{Koretsune2014}, which are adopted in previous theoretical studies~\cite{Naka2019, Naka2020, Misawa2023}.
However, the AFM-induced FFLO superconductivity presented in this paper should be feasible regardless of the details of the parameters.

The model belongs to a wallpaper group $p2gg$ (point group $C_{2v}$)~\cite{Hayami2020} that is composed of translations, twofold rotation $C_{2z}$, and glide symmetries $G_x$ and $G_y$ with respect to the $x$ and $y$ axes, respectively~%
\footnote{In the actual three-dimensional $\kappa$-Cl, whose space group is $Pnma$, the glide operation $G_y$ changes to the $n$ glide with an \textit{out-of-plane} half translation as well as an in-plane one. However, this property does not influence the conclusion of this paper.}:
\begin{subequations}
 \label{eq:symmetry_relation}
 \begin{align}
  \hat{U}_{C_{2z}}^\dagger \hat{H}_0(\bm{k}) \hat{U}_{C_{2z}} &= \hat{H}_0(-\bm{k}), \\
  \hat{U}_{G_x}(\bm{k})^\dagger \hat{H}_0(\bm{k}) \hat{U}_{G_x}(\bm{k}) &= \hat{H}_0(-k_x, k_y), \\
  \hat{U}_{G_y}(\bm{k})^\dagger \hat{H}_0(\bm{k}) \hat{U}_{G_y}(\bm{k}) &= \hat{H}_0(k_x, -k_y).
 \end{align}
\end{subequations}
Here the unitary matrices representing the symmetry operations are defined by
\begin{subequations}
 \label{eq:symmetry_unitary}
 \begin{align}
  \hat{U}_{C_{2z}} &:=
  \begin{bmatrix}
   0 & 1 & 0 & 0 \\
   1 & 0 & 0 & 0 \\
   0 & 0 & 0 & 1 \\
   0 & 0 & 1 & 0
  \end{bmatrix}
  \otimes (-\ii\hat{\sigma}_z), \displaybreak[2] \\
  \hat{U}_{G_x}(\bm{k}) &:=
  \begin{bmatrix}
   0 & 0 & 0 & \ee^{-\ii k_x} \\
   0 & 0 & \ee^{\ii k_y} & 0 \\
   0 & 1 & 0 & 0 \\
   \ee^{\ii (- k_x + k_y)} & 0 & 0 & 0
  \end{bmatrix}
  \otimes (-\ii\hat{\sigma}_x), \displaybreak[2] \\
  \hat{U}_{G_y}(\bm{k}) &:=
  \begin{bmatrix}
   0 & 0 & 1 & 0 \\
   0 & 0 & 0 & \ee^{\ii (k_x + k_y)} \\
   \ee^{\ii k_x} & 0 & 0 & 0 \\
   0 & \ee^{\ii k_y} & 0 & 0
  \end{bmatrix}
  \otimes (-\ii\hat{\sigma}_y).
 \end{align}
\end{subequations}
Note that all the symmetry properties in Eqs.~\eqref{eq:symmetry_relation} are conserved even for a nonzero AFM molecular field $h$, since the unitary matrices in Eqs.~\eqref{eq:symmetry_unitary} include the rotation in the spin space.
If we do not consider the spin rotations (i.e., the spin part is equal to $\hat{\sigma}_0$), the glide symmetries $G_x$ and $G_y$ are broken in the AFM state ($h \neq 0$)~\cite{Naka2019, Hayami2020}.
On the other hand, the time-reversal symmetry $T$, which is preserved in the paramagnetic phase ($h = 0$), is broken in the AFM phase ($h \neq 0$):
\begin{align}
 \hat{U}_T^\dagger \hat{H}_0(\bm{k})^* \hat{U}_T &= \hat{H}_0(-\bm{k}) \ \text{for $h = 0$}, \\
 \hat{U}_T &:= \hat{\bm{1}}_4 \otimes \ii\hat{\sigma}_y,
\end{align}
where $\hat{\bm{1}}_4$ represents the $4 \times 4$ identity matrix.

The noninteracting Hamiltonian $\hat{H}_0(\bm{k})$ is diagonalized using a unitary matrix $\hat{V}_{\text{band}}(\bm{k})$ as
\begin{align}
 & \hat{V}_{\text{band}}(\bm{k})^\dagger \hat{H}_0(\bm{k}) \hat{V}_{\text{band}}(\bm{k}) \notag \\
 &= \diag[\varepsilon(\bm{k})_{1 \uparrow}, \varepsilon(\bm{k})_{1 \downarrow}, \dots, \varepsilon(\bm{k})_{4 \uparrow}, \varepsilon(\bm{k})_{4 \downarrow}],
 \label{eq:H0_diagonalization}
\end{align}
where $\varepsilon(\bm{k})_{\alpha s}$ is the $\alpha$th eigenenergy for spin $s$~%
\footnote{Note that $s$ represents the \textit{pure} spin degree of freedom since the spin--orbit coupling is not included in Eq.~\eqref{eq:Hamiltonian_momentum}}.
Figures~\ref{fig:lattice_electronic_structure}(b) and \ref{fig:lattice_electronic_structure}(c) show the energy band dispersions $\varepsilon(\bm{k})_{\alpha s}$ for $h = 0$ and $h = 0.2$ eV, respectively.
When the AFM molecular field $h$ is finite, the spin splitting appears at general $\bm{k}$ points, except along the $k_{x, y}$ axes and the Brillouin zone boundary~\cite{Naka2019}, as shown in Fig.~\ref{fig:lattice_electronic_structure}(c).
Most of the $\kappa$-type salts possess three electrons per dimer on average, and then the electron density $n = n_\uparrow + n_\downarrow$ per ET molecule is $1.5$.
In this case, by increasing $h$ the system turns at $h \approx 0.14$ eV to an insulating state with finite energy gap [Fig.~\ref{fig:lattice_electronic_structure}(c)].
In the following discussions, we focus on the metallic region for $h \leq 0.14$ eV in the undoped case.
Furthermore, we investigate the effect of carrier doping by changing the electron density $n$ from a hole-doped regime ($n = 1.3$) to an electron-doped regime ($n = 1.7$) [see the green dashed lines in Figs.~\ref{fig:lattice_electronic_structure}(b) and \ref{fig:lattice_electronic_structure}(c)].
In the doped regimes, we consider the wider range $h \leq 0.2$ eV of the AFM molecular field in the metallic regime.

As mentioned above, here we choose the expressions of the Hamiltonian [Eq.~\eqref{eq:Hamiltonian_momentum_kin}] and the symmetry operations [Eqs.~\eqref{eq:symmetry_unitary}] without the phase factors concerning internal coordinates of the molecules, which are irrelevant to the one-particle properties.
In the case of susceptibilities, however, a careful treatment on the phase factors must be conducted, as we will discuss later (see Sec.~\ref{sec:susceptibility} and Appendix~\ref{app:susceptibility_sym}).

In the following, to discuss SC states, we consider interaction effects in addition to the noninteracting tight-binding Hamiltonian introduced above.
In Secs.~\ref{sec:effective_model} and \ref{sec:Eliashberg_Hubbard}, an effective theory with intraband attractive interactions and the Eliashberg theory with the repulsive Hubbard interactions, respectively, are discussed.

\section{Analysis of effective model}
\label{sec:effective_model}

\subsection{Attractive intraband interactions}
In this section, we introduce effective attractive interactions to search for possible AFM-induced FFLO states.
As a preparation, we transform the noninteracting Hamiltonian Eq.~\eqref{eq:Hamiltonian} to the band-based representation.
By using the unitary matrix $\hat{V}_{\text{band}}(\bm{k})$ in Eq.~\eqref{eq:H0_diagonalization}, we define the annihilation and creation operators of electrons on the band $\alpha$ as
\begin{subequations}
 \begin{align}
  d_{\bm{k}, \alpha s} &:= \sum_{l} V_{\text{band}}(\bm{k})^*_{l s, \alpha s} c_{\bm{k}, l s}, \\
  d_{\bm{k}, \alpha s}^\dagger &:= \sum_{l} c_{\bm{k}, l s}^\dagger V_{\text{band}}(\bm{k})_{l s, \alpha s}.
 \end{align}
\end{subequations}
Then the Hamiltonian is rewritten in the band-based form
\begin{equation}
 H_0 = \frac{1}{N} \sum_{\bm{k}} \sum_{\alpha, s} \varepsilon(\bm{k})_{\alpha s} d_{\bm{k}, \alpha s}^\dagger d_{\bm{k}, \alpha s}.
\end{equation}
As for the interaction term, we assume the intraband attractive pairing as an ideal situation, which is given by
\begin{equation}
 H_{\text{int}}^{\text{(band)}} = - \frac{1}{N} \sum_{\bm{Q}} \sum_{p, \alpha} U_p B_{p \alpha}(\bm{Q})^\dagger B_{p \alpha}(\bm{Q}),
\end{equation}
where
\begin{equation}
 B_{p \alpha}(\bm{Q})^\dagger = \frac{1}{\sqrt{2N}} \sum_{\bm{k}} \sum_{s, s'} \psi_{p}(\bm{k}) (\ii\hat{\sigma}_y)_{s s'} d_{\bm{k} + \bm{Q}, \alpha s}^\dagger d_{-\bm{k}, \alpha s'}^\dagger
\end{equation}
is the creation operator of spin-singlet Cooper pairs with a COM momentum $\bm{Q}$ on the band $\alpha$, and $\psi_{p}(\bm{k})$ indicates the momentum dependence of the order parameter with different anisotropy in $\bm{k}$ space indexed by $p$.
Although the pairing symmetry in $\kappa$-(ET)$_2X$ is experimentally not fully determined, spin-singlet nodal superconductivity is supported by many measurements~\cite{[{For reviews, see e.g., }][]{Wosnitza1999_review}, *Kuroki2006_review}.
Therefore, we here consider the extended $s + d_{x^2-y^2}$-wave superconductivity in the $A_1$ irreducible representation (irrep) of point group $C_{2v}$, and the $d_{xy}$-wave one belonging to the $A_2$ irrep (see Table~\ref{tab:character_C2v})~%
\footnote{When we consider three-dimensional models containing $z$ axis, the SC order parameter is classified by irreps of other point groups, e.g., $D_{2h}$~\cite{Powell2006}}.
Following the previous studies~\cite{Kuroki2002, Sekine2013, Guterding2016, HiroshiWatanabe2017, HiroshiWatanabe2019}, the basis function $\psi_{p}(\bm{k})$ is chosen as
\begin{subequations}
 \begin{align}
  \psi_1(\bm{k}) &= 2 \cos k_x \cos k_y, \\
  \psi_2(\bm{k}) &= \cos k_x, \\
  \psi_3(\bm{k}) &= \cos k_y, \\
  \intertext{for the extended $s + d_{x^2-y^2}$-wave ($A_1$) order, and}
  \psi_4(\bm{k}) &= 2 \sin k_x \sin k_y,
 \end{align}
\end{subequations}
for the $d_{xy}$-wave ($A_2$) order.
In the following calculations, we assume two kinds of interaction parameters,
\begin{equation}
 (U_1, U_2, U_3, U_4) =
 \begin{cases}
  (0.1, 0.1, 0.1, 0) & \text{$A_1$ case}, \\
  (0, 0, 0, 0.1) & \text{$A_2$ case}.
 \end{cases}
\end{equation}
Note that the results in Sec.~\ref{sec:effective_model_sus} are not qualitatively altered by the choice of the interaction parameters, as long as the transition temperature $T_c$ is much smaller than the Fermi energy.

\begin{table}[tbp]
 \caption{Character table for the wallpaper group $p2gg$ (point group $C_{2v}$). $E$ represents the identity operation.}
 \label{tab:character_C2v}
 \begin{tabular}{cccccc} \hline\hline
  Irrep & $E$ & $C_{2z}$ & $G_x$ & $G_y$ & Basis functions \\ \hline
  $A_1$ & $1$ & $1$ & $1$ & $1$ & $1$, $k_x^2 - k_y^2$ \\
  $A_2$ & $1$ & $1$ & $-1$ & $-1$ & $k_x k_y$ \\
  $B_1$ & $1$ & $-1$ & $-1$ & $1$ & $k_y \hat{z}$ \\
  $B_2$ & $1$ & $-1$ & $1$ & $-1$ & $k_x \hat{z}$ \\ \hline\hline
 \end{tabular}
\end{table}

\begin{figure*}
 \includegraphics[width=\linewidth, pagebox=artbox]{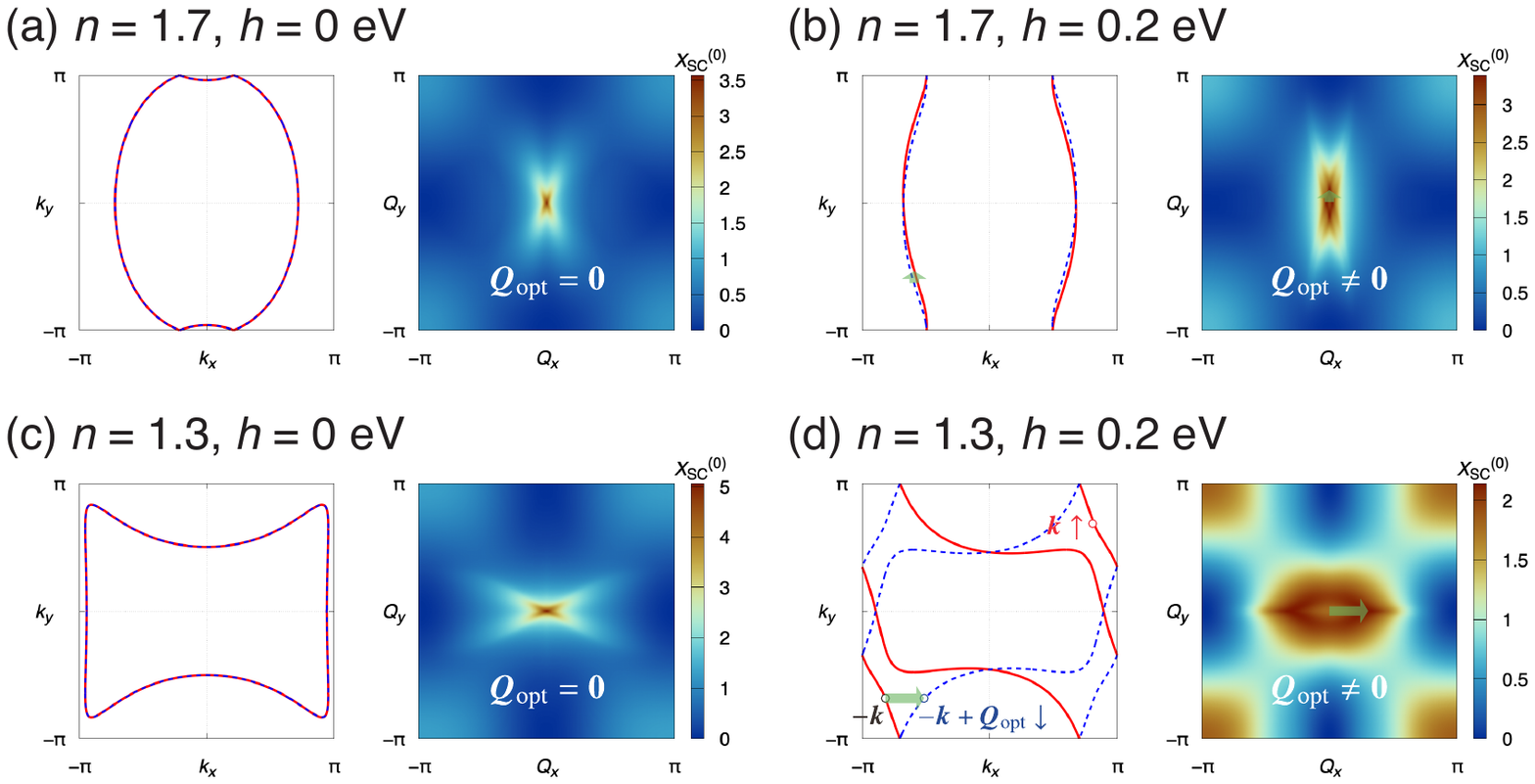}
 \caption{Fermi surfaces and $X_{\text{SC}}^{(0)}(\bm{Q})$ for (a) $(n, h) = (1.7, 0)$, (b) $(n, h) = (1.7, 0.2)$, (c) $(n, h) = (1.3, 0)$, and (d) $(n, h) = (1.3, 0.2)$. The temperature $T$ is set to 1 meV. The red solid (blue dashed) lines represent Fermi surfaces of the up-spin (down-spin) electrons, which are degenerate for $h = 0$. The $A_2$-type ($d_{xy}$-wave) interaction parameters are used in the calculations of $X_{\text{SC}}^{(0)}(\bm{Q})$ for all cases. In the paramagnetic phase [(a), (c)], the maximum of $X_{\text{SC}}^{(0)}(\bm{Q})$ is located at $\bm{Q} = \bm{0}$. In the AFM phase [(b), (d)], $X_{\text{SC}}^{(0)}(\bm{Q})$ has peaks at finite momenta indicating the FFLO state.}
 \label{fig:fs_susSC}
\end{figure*}

\subsection{Linearized gap equation}
\label{sec:effective_model_sus}
Considering the Hamiltonian $H = H_0 + H_{\text{int}}^{\text{(band)}}$, we now study whether the FFLO state is realized through the analysis of a linearized gap equation, and clarify the SC instability just above $T_c$.
The linearized gap equation is formulated by calculating the SC susceptibility matrix whose matrix elements are defined by
\begin{align}
 & \chi_{\text{SC}}(\bm{Q}, \ii\omega_n)_{p\alpha, p'\alpha'} \notag \\
 &:= \int_{0}^{\beta} \mathrm{d}\tau \, \ee^{\ii\omega_n\tau} \Braket{B_{p \alpha}(\bm{Q}, \tau) B_{p' \alpha'}(\bm{Q}, 0)^\dagger},
\end{align}
where $\omega_n = 2n\pi T$ is the bosonic Matsubara frequency, $\beta = 1/T$ is the inverse temperature, and $B_{p \alpha}(\bm{Q}, \tau) = \ee^{\tau H} B_{p \alpha}(\bm{Q}) \ee^{-\tau H}$.
The SC susceptibility is obtained by using the $T$-matrix approximation as
\begin{align}
 \hat{\chi}_{\text{SC}}(Q) &= \left[ \hat{\bm{1}} - \hat{\chi}_{\text{SC}}^{(0)}(Q) \hat{U}^{\text{(band)}} \right]^{-1} \hat{\chi}_{\text{SC}}^{(0)}(Q), \\
 U^{\text{(band)}}_{p\alpha, p'\alpha'} &:= U_p \delta_{p, p'} \delta_{\alpha, \alpha'},
\end{align}
where $Q$ stands for $(\bm{Q}, \ii\omega_n)$, and $\hat{\bm{1}}$ represents the identity matrix.
The irreducible susceptibility is given by
\begin{align}
 \chi_{\text{SC}}^{(0)}(Q)_{p\alpha, p'\alpha'}
 &= \frac{T \delta_{\alpha, \alpha'}}{2N} \sum_{k} \sum_{s} \tilde{G}^{(0)}(k+Q)_{\alpha s} \tilde{G}^{(0)}(-k)_{\alpha \bar{s}} \notag \\
 &\qquad \times \psi_p(\bm{k})^* [\psi_{p'}(\bm{k}) + \psi_{p'}(-\bm{k}-\bm{Q})],
 \label{eq:susSC0}
\end{align}
where
\begin{equation}
 \tilde{G}^{(0)}(k)_{\alpha s} = \tilde{G}^{(0)}(\bm{k}, \ii\varepsilon_m)_{\alpha s}
 = \frac{1}{\ii\varepsilon_m + \mu_0 - \varepsilon(\bm{k})_{\alpha s}}
\end{equation}
is the noninteracting band-based Green's function~%
\footnote{The noninteracting Green's function has no off-diagonal component in the band basis due to the absence of the spin--orbit coupling.}
and $\varepsilon_m = (2m + 1)\pi T$ is the fermionic Matsubara frequency.
$\mu_0$ is the chemical potential for the noninteracting Hamiltonian.
Since the SC transition occurs when $\hat{\chi}_{\text{SC}}(Q)$ diverges, the criterion of the SC instability is obtained when the largest eigenvalue of $\hat{\chi}_{\text{SC}}^{(0)}(Q) \hat{U}^{\text{(band)}}$ becomes unity.
We assume that the bosonic Matsubara frequency $\omega_n$ is always zero for the criterion.
In this case, the irreducible susceptibility is simplified by performing the summation of the frequency in Eq.~\eqref{eq:susSC0},
\begin{align}
 & \chi_{\text{SC}}^{(0)}(\bm{Q}, 0)_{p\alpha, p'\alpha'} \notag \\
 &= - \frac{\delta_{\alpha, \alpha'}}{2N} \sum_{\bm{k}} \sum_s \frac{f[\varepsilon(\bm{k}+\bm{Q})_{\alpha s}] - f[-\varepsilon(-\bm{k})_{\alpha \bar{s}}]}{\varepsilon(\bm{k}+\bm{Q})_{\alpha s} + \varepsilon(-\bm{k})_{\alpha \bar{s}}} \notag \\
 &\qquad \times \psi_{p}(\bm{k})^* [\psi_{p'}(\bm{k}) + \psi_{p'}(-\bm{k}-\bm{Q})],
\end{align}
where $f(\varepsilon) = (\ee^{\beta (\varepsilon - \mu_0)} + 1)^{-1}$ is the Fermi--Dirac distribution function.

Let $X_{\text{SC}}^{(0)}(\bm{Q})$ be the largest eigenvalue of $\hat{\chi}_{\text{SC}}^{(0)}(\bm{Q}, 0) \hat{U}^{\text{(band)}}$ for a fixed $\bm{Q}$.
We then find the optimal COM momentum(s) $\bm{Q}_{\text{opt}}$ of the Cooper pairs near $T_c$ such that the eigenvalue has a maximum value as
\begin{equation}
 X_{\text{SC}}^{(0)}(\bm{Q}_{\text{opt}}) = \max_{\bm{Q}} X_{\text{SC}}^{(0)}(\bm{Q}).
\end{equation}
For example, we show the structure of $X_{\text{SC}}^{(0)}(\bm{Q})$ for the $d_{xy}$-wave ($A_2$) order parameter in the electron-doped ($n = 1.7$) and hole-doped ($n = 1.3$) cases in Figs.~\ref{fig:fs_susSC}(a)--\ref{fig:fs_susSC}(b) and \ref{fig:fs_susSC}(c)--\ref{fig:fs_susSC}(d), respectively [for the undoped ($n = 1.5$) situation, see Appendix~\ref{app:undoped}].
When $h = 0$, the whole Fermi surfaces are spin degenerate, and $X_{\text{SC}}^{(0)}(\bm{Q})$ reaches a maximum at $\bm{Q} = \bm{0}$ [Figs.~\ref{fig:fs_susSC}(a) and \ref{fig:fs_susSC}(c)]; this indicates the SC state with zero COM momentum of the Cooper pairs.

In the presence of the AFM field $h$, on the other hand, the optimal COM momentum $\bm{Q}_{\text{opt}}$ becomes nonzero because of the spin splitting [Figs.~\ref{fig:fs_susSC}(b) and \ref{fig:fs_susSC}(d)].
This can be understood as follows: when a spin-up state is located at a general $\bm{k}$ point on the Fermi surfaces, there is only a spin-up state at $- \bm{k}$ for the $T$-breaking and $C_{2z}$-preserving AFM order, whereas the equal-spin pairing is prohibited in the intraband $d$-wave superconductivity.
Instead, to stabilize the spin-singlet superconductivity, a finite momentum shift $\bm{Q}$ is necessary for the pairing between the spin-up $\bm{k}$ and spin-down $- \bm{k} + \bm{Q}$ states [see Fig.~\ref{fig:fs_susSC}(d)].
This mechanism is similar to the FFLO state under Zeeman field.

One can see in Fig.~\ref{fig:fs_susSC} that the magnitudes of the splitting are quite different between the electron- and hole-doped cases; the Fermi surfaces and their splitting are quasi-one-dimensional and small (two-dimensional and large) in the electron-doped (hole-doped) regime.
As a result, $\bm{Q}_{\text{opt}}$ in the hole-doped regime tends to be larger than that in the electron-doped regime.
Also, reflecting the difference in the structures of the Fermi surfaces, the direction of $\bm{Q}_{\text{opt}}$ is parallel to $y$ and $x$ axes in the electron- and hole-doped cases, respectively.
Furthermore, $X_{\text{SC}}^{(0)}(\bm{Q}) = X_{\text{SC}}^{(0)}(-\bm{Q})$ is always satisfied because of the presence of the twofold rotation symmetry (or inversion symmetry), which results in the double-peak structure at $\pm\bm{Q}_{\text{opt}}$ of $X_{\text{SC}}^{(0)}(\bm{Q})$.
We will revisit this point in Sec.~\ref{sec:discussion}.

\begin{figure}
 \includegraphics[width=\linewidth, pagebox=artbox]{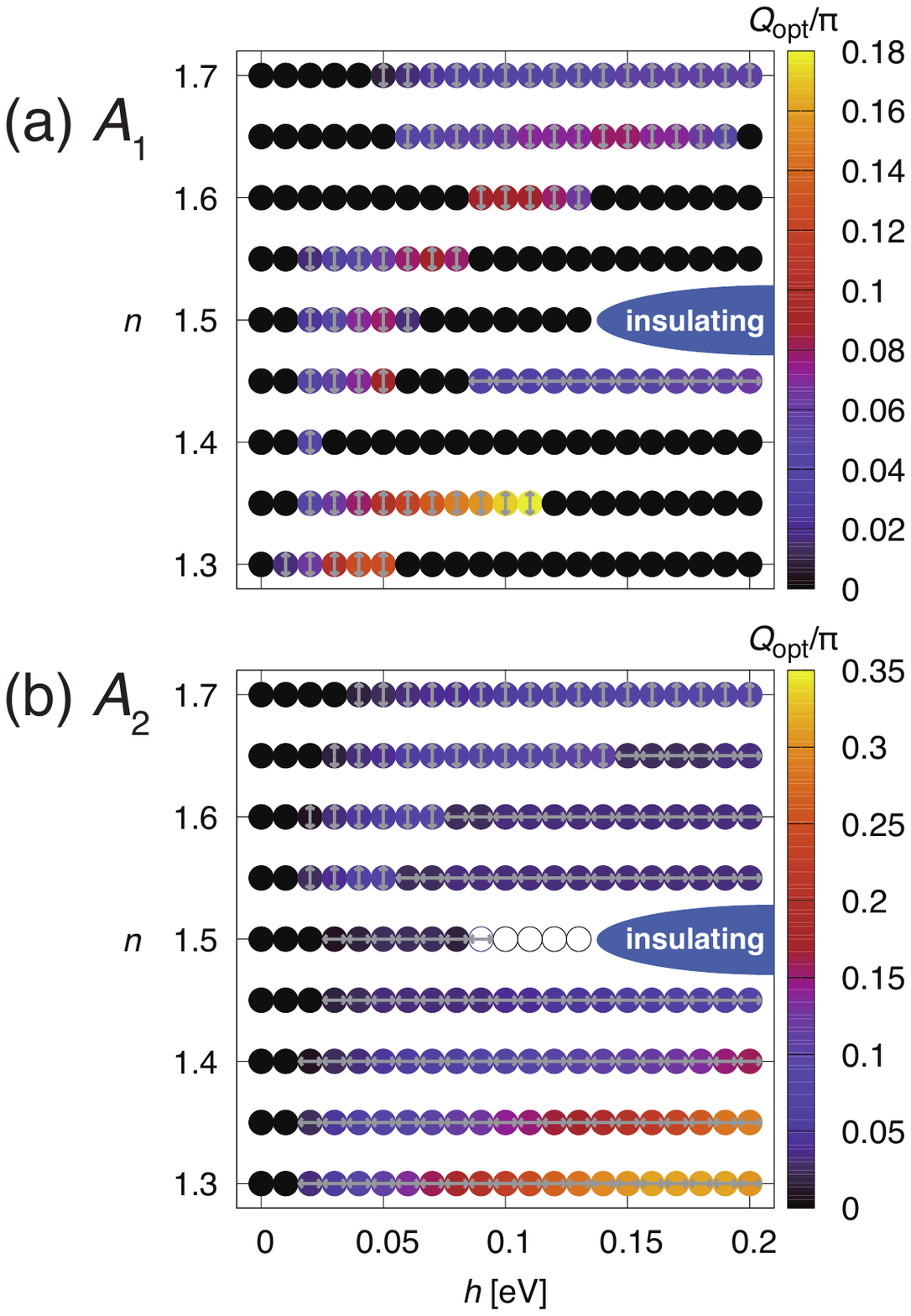}
 \caption{Phase diagrams obtained from the analysis of the SC susceptibility for (a) $A_1$ and (b) $A_2$ interaction parameters. The temperature $T$ is set to 1 meV. The closed (open) circles indicate that the largest eigenvalue of $X_{\text{SC}}^{(0)}(\bm{Q})$ is larger (smaller) than unity. The color of each circle represents the norm of the optimal COM momentum $Q_{\text{opt}} = |\bm{Q}_{\text{opt}}|$. The direction of $\bm{Q}_{\text{opt}}$ is represented by gray arrows in the circles.}
 \label{fig:phase_diagram_effective}
\end{figure}

In this way, we analyze $X_{\text{SC}}^{(0)}(\bm{Q})$ by varying the electron density $n$ and the AFM molecular field $h$.
The results for the extended $s + d_{x^2-y^2}$-wave ($A_1$) and $d_{xy}$-wave ($A_2$) superconductivity are shown in Figs.~\ref{fig:phase_diagram_effective}(a) and \ref{fig:phase_diagram_effective}(b), respectively.
Both phase diagrams contain a parameter region with $h > 0$, where $X_{\text{SC}}^{(0)}(\bm{Q})$ has peaks at finite momenta $\pm \bm{Q}_{\text{opt}}$ (the colored circles), indicating the appearance of the AFM-induced FFLO state.
The finite COM state is obtained in a broad parameter region for the $A_2$-type interaction, whereas it is restricted to a narrower regime in the $A_1$-type case.
The reason is intuitively understood as follows, considering the gap structure for the two cases~\cite{Kuroki2002}.
When the interaction is $A_2$-type, the $d_{xy}$-wave SC order parameter is zero on the $k_{x, y}$ axes and the Brillouin zone boundary, where the spin degeneracy is protected even in the AFM state~\cite{Naka2019}.
Therefore, the $A_2$ superconductivity is more susceptible to the presence of the spin splitting; the finite COM momentum is necessary to form the spin-singlet Cooper pairs on the Fermi surfaces even when the splitting is small.
On the other hand, the $A_1$ case is not so sensitive to the spin splitting, since zeros of the extended $s + d_{x^2-y^2}$-wave order parameter are in general $\bm{k}$ points and therefore do not necessarily correspond to the spin-degenerate regions.

\section{Eliashberg theory based on Hubbard model}
\label{sec:Eliashberg_Hubbard}
In the previous section, we have demonstrated that the FFLO phase can be stabilized due to the spin splitting in the AFM state by using the effective model with attractive pairing interactions.
In this section, we consider the Eliashberg theory based on the repulsive Hubbard model~\cite{Kino1996} as a more realistic description.
The Hubbard Hamiltonian is written as $H = H_0 + H_{\text{int}}^{\text{(Hub)}}$, where the interaction term is given by
\begin{equation}
 H_{\text{int}}^{\text{(Hub)}} = U \sum_{\bm{R}} \sum_{l} n_{\bm{R}, l\uparrow} n_{\bm{R}, l\downarrow},
 \label{eq:interaction_Hubbard}
\end{equation}
$U$ being the on-site Coulomb repulsion on the ET molecule.
$n_{\bm{R}, l s} = c_{\bm{R}, l s}^\dagger c_{\bm{R}, l s}$ is the electron-number operator; we remind that $\bm{R}$, $l$, and $s$ represent the unit cell position, the sublattice index and spin, respectively.
Equation~\eqref{eq:interaction_Hubbard} is then rewritten as
\begin{equation}
 H_{\text{int}}^{\text{(Hub)}} = \frac{1}{4} \sum_{\bm{R}} \sum_{\zeta_1, \dots, \zeta_4} U^{\text{(Hub)}}_{\zeta_1 \zeta_2, \zeta_3 \zeta_4} c_{\bm{R}, \zeta_1}^\dagger c_{\bm{R}, \zeta_2} c_{\bm{R}, \zeta_3} c_{\bm{R}, \zeta_4}^\dagger,
\end{equation}
where $\zeta_i$ stands for $(l_i, s_i)$, and
\begin{gather}
 U^{\text{(Hub)}}_{\zeta_1 \zeta_2, \zeta_3 \zeta_4} = \delta_{l_1, l_2} \delta_{l_2, l_3} \delta_{l_3, l_4} U_{s_1 s_2, s_3 s_4}, \\
 U_{s_1 s_2, s_3 s_4} =
 \begin{cases}
  U & (s_1 s_2, s_3 s_4) = (\uparrow \downarrow, \uparrow \downarrow) \ \text{or} \  (\downarrow \uparrow, \downarrow \uparrow), \\
  -U & (s_1 s_2, s_3 s_4) = (\uparrow \uparrow, \downarrow \downarrow) \ \text{or} \  (\downarrow \downarrow, \uparrow \uparrow), \\
  0 & \text{otherwise}.
 \end{cases}
\end{gather}
For the later formulations, we define a matrix form of the Hubbard interaction: $\hat{U}^{\text{(Hub)}} := [U^{\text{(Hub)}}_{\zeta_1 \zeta_2, \zeta_3 \zeta_4}]$.
In the following Secs.~\ref{sec:susceptibility} and \ref{sec:linearized_Eliashberg}, a generalized susceptibility based on the FLEX approximation and a linearized Eliashberg equation are discussed, respectively.
The source code used for the numerical calculations is available in Ref.~\cite{SourceCode}, some parts of which are implemented based on the algorithm of the FLEX$+$IR package~\cite{Witt2021}.

Before going into detail, we make several comments on the microscopic analyses in the $\kappa$-type salts.
First, the correlation effects included in the FLEX approximation are important for the emergence of superconductivity.
Indeed, our test calculations within the random phase approximation (RPA) indicate that the divergence of the magnetic susceptibility occurs far before the SC transition.
Second, we emphasize the differences from previous related studies~\cite{Kuroki2002, Sekine2013}.
They have discussed extended $s + d_{x^2-y^2}$-wave ($A_1$) versus $d_{xy}$-wave ($A_2$) pairing instability in $\kappa$-type superconductors by using the FLEX approximation and the Eliashberg theory, but considered only the undoped ($n = 1.5$), paramagnetic ($h = 0$), and zero-COM-momentum ($\bm{Q} = \bm{0}$) cases.
In the present paper, we consider finite doping $n$, extend to the presence of AFM field $h$, and seek the possibility of a finite COM momentum $\bm{Q}$.

\subsection{Generalized susceptibility}
\label{sec:susceptibility}
Now let us formulate the generalized susceptibility within the FLEX approximation.
The noninteracting Green's function for $U = 0$ is represented by the eight-dimensional (four sublattices $\times$ two directions of spin) matrix:
\begin{equation}
 \hat{G}^{(0)}(\bm{k}, \ii\varepsilon_m) = \left[ (\ii\varepsilon_m + \mu_0) \hat{\bm{1}} - \hat{H}_0(\bm{k}) \right]^{-1}.
\end{equation}
In the interacting case ($U \neq 0$), the dressed Green's function is given by
\begin{equation}
 \hat{G}(\bm{k}, \ii\varepsilon_m) = \left[ (\ii\varepsilon_m + \mu) \hat{\bm{1}} - \hat{H}_0(\bm{k}) - \hat{\Sigma}(\bm{k}, \ii\varepsilon_m) \right]^{-1},
 \label{eq:Green_func_normal}
\end{equation}
where $\mu$ and $\hat{\Sigma}(\bm{k}, \ii\varepsilon_m)$ are the chemical potential self-consistently determined in the interacting system and the (normal) self-energy, respectively; the matrix elements of the self-energy are given by
\begin{equation}
 \Sigma(k)_{\zeta, \zeta'} = \frac{T}{N} \sum_{q} \sum_{\zeta_1, \zeta_2} V^{\text{(n)}}(q)_{\zeta \zeta_1, \zeta' \zeta_2} G(k-q)_{\zeta_1, \zeta_2},
\end{equation}
with $k$ and $q$ representing $(\bm{k}, \ii\varepsilon_m)$ and $(\bm{q}, \ii\omega_n)$, respectively.
Within the FLEX approximation, the effective interaction vertex for the normal part is calculated as
\begin{equation}
 \hat{V}^{\text{(n)}}(q) = \hat{U}^{\text{(Hub)}} \left[ \hat{\chi}(q) - \frac{1}{2} \hat{\chi}^{(0)}(q) \right] \hat{U}^{\text{(Hub)}},
\end{equation}
where
\begin{align}
 \chi^{(0)}(q)_{\zeta_1 \zeta_2, \zeta_3 \zeta_4} &= - \frac{T}{N} \sum_{k} G(k+q)_{\zeta_1, \zeta_3} G(k)_{\zeta_4, \zeta_2}, \\
 \hat{\chi}(q) &= \left[ \hat{\bm{1}} - \hat{\chi}^{(0)}(q) \hat{U}^{\text{(Hub)}} \right]^{-1} \hat{\chi}^{(0)}(q),
 \label{eq:generalized_susceptibility}
\end{align}
are the bare and generalized susceptibilities, respectively.
In the numerical study, we take $32 \times 32$ $\bm{k}$-point meshes and about 80 Matsubara frequencies generated by the \texttt{SparseIR.jl} package~\cite{Wallerberger2023, Wang2020} based on the intermediate representation~\cite{Shinaoka2017} and the sparse sampling~\cite{Li2020}.
The temperature $T$ and the energy cutoff $\omega_{\text{max}}$ are set to 1 meV and 2 eV, respectively.

We show the results for $U = 1$ eV in the following.
In general, the FLEX approximation is justified within the intermediate-coupling region in which $U$ is smaller than half of the bandwidth $W$, namely, $U / W \lesssim 0.5$.
Although our theory adopts the interaction comparable to the bandwidth ($U \sim W$), in the three-quarter-filled dimerized system that we treat, the effective \textit{intradimer} Coulomb interaction, which acts on the two electrons in the antibonding orbital in the dimer, is estimated as~\cite{Kino1996, Naka2010, Naka2013}
\begin{equation}
 U_{\text{eff}} = \frac{U}{2} + 2|t_a| - \sqrt{\left(\frac{U}{2}\right)^2 + (2t_a)^2} \approx 0.265 \ \text{eV},
\end{equation}
which is approximately equal to $W / 4$.
Therefore, our choice of $U$ does not cause a serious problem even in the FLEX approximation.

Using the generalized susceptibility [Eq.~\eqref{eq:generalized_susceptibility}], the dynamical susceptibility for any multipole operator $\hat{\mathcal{O}}$ is generally given by
\begin{equation}
 \tilde{\chi}_{\hat{\mathcal{O}}}(q) = \sum_{\zeta_1, \dots, \zeta_4} \ee^{-\ii \bm{q} \cdot (\bm{r}_{l_1} - \bm{r}_{l_3})} \mathcal{O}_{\zeta_1, \zeta_2} \chi(q)_{\zeta_1 \zeta_2, \zeta_3 \zeta_4} \mathcal{O}_{\zeta_3, \zeta_4},
 \label{eq:multipole_susceptibility}
\end{equation}
where $\bm{r}_l$ is the internal coordinate of the sublattice $l$, namely, the relative position from the origin of the unit cell.
The phase factor $\ee^{-\ii \bm{q} \cdot (\bm{r}_{l_1} - \bm{r}_{l_3})}$ in the summation is necessary to recover the glide symmetries ($G_x$ and $G_y$) of the susceptibility (see Appendix~\ref{app:susceptibility_sym} for details).
The multipole $\hat{\mathcal{O}}$ can be represented by the Kronecker product of a $4 \times 4$ matrix $\hat{M}_{\text{sl}}$ in the sublattice space and a $2 \times 2$ matrix $\hat{M}_{\text{sp}}$ in the spin space.
Although the former sublattice part is classified into 16 types of matrices using the cluster multipole description~\cite{Hayami2020}, we here focus on two representatives: the ferroic matrix $\hat{\bm{1}}_4$ and the antiferroic one $\hat{M}_{\text{AF}}$.
Table~\ref{tab:classification_multipole} shows the classification of the operator $\hat{\mathcal{O}}$ in the two cases.
We calculate Eq.~\eqref{eq:multipole_susceptibility} for all operators in the table, and confirm that the susceptibility for the longitudinal AFM (LAFM) or transverse AFM (TAFM) spin operator has the leading contribution, which are degenerate for $h = 0$.
Whether the LAFM or TAFM susceptibility is dominant for $h > 0$ depends on the parameter choice (see Appendix~\ref{app:LAFM_TAFM} for details).

\begin{table}
 \begin{center}
  \caption{Classification of the multipole operator $\hat{\mathcal{O}} = (\hat{M}_{\text{sl}} \otimes \hat{M}_{\text{sp}}) / 2\sqrt{2}$, where the coefficient $1 / 2\sqrt{2}$ is introduced to satisfy the normalization condition $\tr[\hat{\mathcal{O}}^\dagger \hat{\mathcal{O}}] = 1$. $\hat{\sigma}_\pm = (\hat{\sigma}_x \pm \ii\hat{\sigma}_y) / {\sqrt{2}}$ are ladder operators for spin.}
  \label{tab:classification_multipole}
  \begin{tabular}{ccc} \hline\hline
   \backslashbox{$\hat{M}_{\text{sp}}$}{$\hat{M}_{\text{sl}}$} & $\hat{\bm{1}}_4$ & $\hat{M}_{\text{AF}}$ \\ \hline
   $\hat{\sigma}_0$ & Electric charge & Electric quadrupole \\
   $\hat{\sigma}_z$ & Longitudinal FM spin & Longitudinal AFM spin \\
   $\hat{\sigma}_\pm$ & Transverse FM spin & Transverse AFM spin \\ \hline\hline
  \end{tabular}
 \end{center}
\end{table}

\begin{figure*}
 \includegraphics[width=\linewidth, pagebox=artbox]{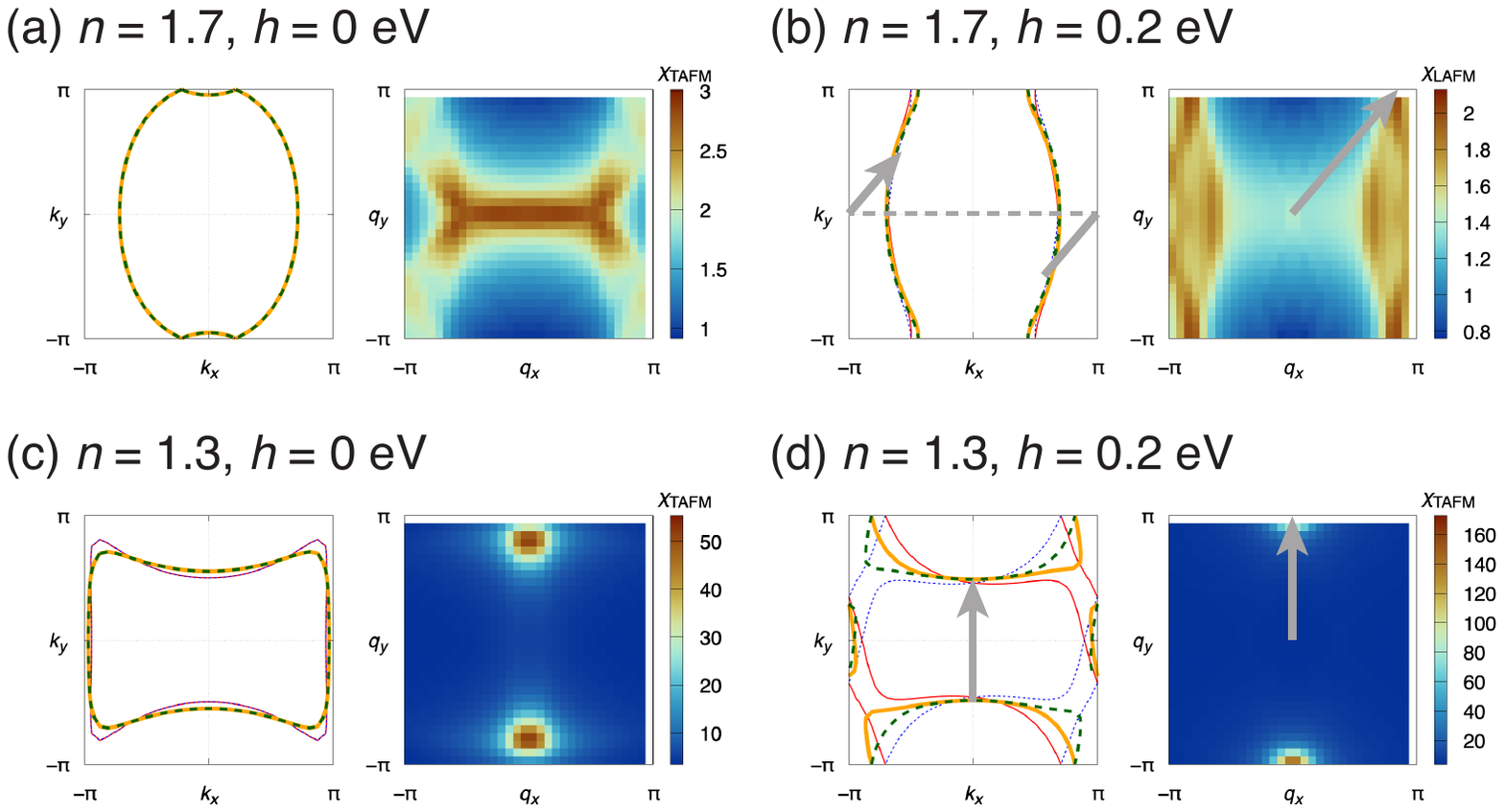}
 \caption{The renormalized Fermi surfaces and $\chi_{\text{LAFM}}(\bm{q})$ or $\chi_{\text{TAFM}}(\bm{q})$ for (a) $(n, h) = (1.7, 0)$, (b) $(n, h) = (1.7, 0.2)$, (c) $(n, h) = (1.3, 0)$, and (d) $(n, h) = (1.3, 0.2)$. The orange solid (green dashed) lines represent the Fermi surfaces of the up-spin (down-spin) electrons taking into account the self-energy, while the red solid (blue dashed) lines show the up-spin (down-spin) noninteracting Fermi surfaces. The gray arrows in (b) and (d) are the $\bm{q}$ vector where the susceptibility reaches maximum, which corresponds to the nesting of the Fermi surfaces.}
 \label{fig:fs_susspin}
\end{figure*}

Here we discuss the relation between the Fermi surfaces and the AFM spin susceptibility by changing the AFM molecular field $h$ and electron density $n$, which is directly related to the SC instability presented in the next subsection.
Let us introduce the Hamiltonian taking the correlation effect into account,
\begin{equation}
 \hat{H}_0(\bm{k}) + \hat{\Sigma}^{\text{R}}(\bm{k}, \omega = 0) - \mu \hat{\bm{1}},
 \label{eq:renormalized_Hamiltonian}
\end{equation}
where the retarded self-energy in the static limit is evaluated by an approximation justified at low temperatures as,
\begin{equation}
 \hat{\Sigma}^{\text{R}}(\bm{k}, \omega = 0) \simeq
 \frac{\hat{\Sigma}(\bm{k}, \ii\pi T) + \hat{\Sigma}(\bm{k}, -\ii\pi T)}{2}.
\end{equation}
We calculate the real part of the right eigenvalues of Eq.~\eqref{eq:renormalized_Hamiltonian} that is non-Hermitian in general; the Fermi surfaces renormalized by the correlations are determined by its zeros.
Figure~\ref{fig:fs_susspin} displays the modified Fermi surfaces and LAFM or TAFM spin susceptibility in the static limit,
\begin{equation}
 \chi_{\text{LAFM}}(\bm{q}) := \tilde{\chi}_{\hat{M}_{\text{AF}} \otimes \hat{\sigma}_z}(\bm{q}, 0), \
 \chi_{\text{TAFM}}(\bm{q}) := \tilde{\chi}_{\hat{M}_{\text{AF}} \otimes \hat{\sigma}_\pm}(\bm{q}, 0),
\end{equation}
where the panels (a), (b) and (c), (d) correspond to the electron-doped ($n = 1.7$) and hole-doped ($n = 1.3$) cases, respectively [for the undoped ($n = 1.5$) regime, see Appendix~\ref{app:undoped}].
The AFM susceptibility with the larger maximum value is shown (see Fig.~\ref{fig:sus_graph} in Appendix~\ref{app:LAFM_TAFM}).
In the presence of the AFM order ($h \neq 0$), as can be seen from the gray arrows in Figs.~\ref{fig:fs_susspin}(b) and \ref{fig:fs_susspin}(d), peaks of the spin susceptibility roughly correspond to the nesting vector of the modified Fermi surfaces.
The Fermi surfaces in the hole-doped AFM state are drastically altered by the correlation effects [Fig.~\ref{fig:fs_susspin}(d)], most likely owing to the larger density of states near a van Hove singularity, which enhances the nesting with the vector $\bm{q} \sim (0, \pi)$.

\subsection{Linearized Eliashberg equation}
\label{sec:linearized_Eliashberg}
We then investigate the SC instability within the framework of Eliashberg theory.
Since the possibility of the FFLO state with the finite COM momentum $\bm{Q}$ is of major interest in the present paper, we define the anomalous Green's function with finite $\bm{Q}$ as
\begin{equation}
 F_{\bm{Q}}(k)_{\zeta, \zeta'} := - \int_{0}^{\beta} \mathrm{d}\tau \, \ee^{\ii\varepsilon_m\tau} \Braket{T_\tau \left[ c_{\bm{k}, \zeta}(\tau) c_{-\bm{k}+\bm{Q}, \zeta'} \right]},
 \label{eq:Green_func_anomalous}
\end{equation}
where $c_{\bm{k}, \zeta}(\tau) = \ee^{\tau H} c_{\bm{k}, \zeta} \ee^{-\tau H}$.
Note that the function satisfies the following relation due to the Fermi--Dirac statistics:
\begin{equation}
 F_{\bm{Q}}(\bm{k}, \ii\varepsilon_n)_{\zeta, \zeta'} = - F_{\bm{Q}}(-\bm{k}+\bm{Q}, \ii\varepsilon_n)_{\zeta', \zeta}.
 \label{eq:Green_func_anomalous_relation}
\end{equation}
Using the normal and anomalous Green's functions in Eqs.~\eqref{eq:Green_func_normal} and \eqref{eq:Green_func_anomalous}, we can construct the Dyson--Gor'kov equation with finite $\bm{Q}$.
By linearizing the equation with respect to the anomalous term, the linearized Eliashberg equation is obtained as
\begin{equation}
 \lambda_{\bm{Q}} \Delta_{\bm{Q}}(k)_{\zeta, \zeta'} = \frac{T}{N} \sum_{q} \sum_{\zeta_1, \zeta_2} V^{\text{(a)}}(q)_{\zeta \zeta_1, \zeta_2 \zeta'} F_{\bm{Q}}(k-q)_{\zeta_1, \zeta_2}.
 \label{eq:linearized_Eliashberg}
\end{equation}
The anomalous Green's function and the interaction vertex are calculated as
\begin{align}
 \hat{F}_{\bm{Q}}(k) &= - \hat{G}(k) \hat{\Delta}_{\bm{Q}}(k) \hat{G}(-k+Q)^{\text{T}}, 
 \label{eq:Green_func_anomalous_2} \\
 \hat{V}^{\text{(a)}}(q) &= - \frac{1}{2} \hat{U}^{\text{(Hub)}} - \hat{U}^{\text{(Hub)}} \hat{\chi}(q) \hat{U}^{\text{(Hub)}},
\end{align}
where $\hat{\Delta}_{\bm{Q}}(k)$ is the order parameter matrix and $Q := (\bm{Q}, 0)$.
We here assume that the interaction vertex $\hat{V}^{\text{(a)}}$ is independent of the COM momentum $\bm{Q}$.
A similar formalism for the finite-$\bm{Q}$ Eliashberg theory has been discussed in a previous study~\cite{Yoshida2021}.

\begin{figure*}
 \includegraphics[width=\linewidth, pagebox=artbox]{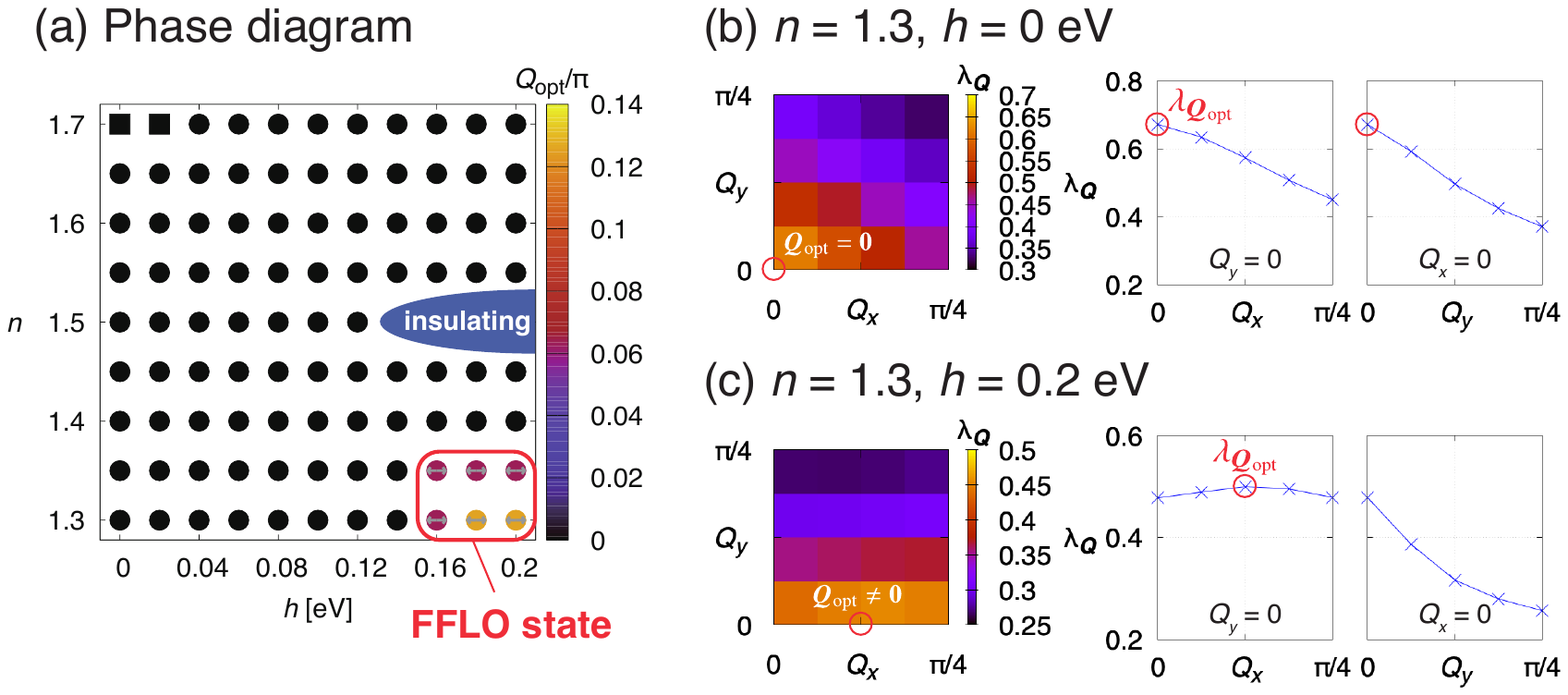}
 \caption{(a) Phase diagram on the $(h, n)$ plane obtained from the analysis of the Hubbard model on the $\kappa$-type structure. The temperature $T$ is set to 1 meV. The squares (circles) indicate that the stable order parameter belongs to the $A_1$ ($A_2$) irrep. The color of each point represents the norm of the optimal COM momentum $Q_{\text{opt}} = |\bm{Q}_{\text{opt}}|$. (b), (c) The momentum dependence of the eigenvalue $\lambda_{\bm{Q}}$ of the Eliashberg equation in the hole-doped regime ($n = 1.3$) for $h = 0$ and $h = 0.2$, respectively. The eigenvalue has peaks at finite COM momenta in (c).}
 \label{fig:phase_diagram_Hubbard}
\end{figure*}

Within the Eliashberg theory, the phase transition into the SC state with the COM momentum $\bm{Q}$ takes place when the eigenvalue $\lambda_{\bm{Q}}$ in Eq.~\eqref{eq:linearized_Eliashberg} reaches unity; the $\bm{Q}$ value whose transition temperature is highest is realized.
Here, for numerical convenience, we evaluate the amplitude of the Eliashberg eigenvalue at a fixed temperature ($T = 1$ meV), which is far below the Neel temperature in $\kappa$-Cl, and judge which $\bm{Q}$ is favored for each $(n, h)$.
By using the power method, we solve the linearized Eliashberg Eq.~\eqref{eq:linearized_Eliashberg} and find the eigenvalue $\lambda_{\bm{Q}}$ and the corresponding order parameter $\hat{\Delta}_{\bm{Q}}(k)$.
As is the analysis in Sec.~\ref{sec:effective_model_sus}, we consider the SC order parameter belonging to the $A_1$ and $A_2$ irreps of $C_{2v}$.
The initial functional form is chosen as
\begin{align}
 \hat{\Delta}_{\bm{Q}}^{\text{(init)}}(k) &\sim
 (\cos\tilde{k}_x - \cos\tilde{k}_y) (\hat{\bm{1}}_4 \otimes \ii\hat{\sigma}_y) \notag \\
 &\quad + \delta \bigl[ \sin\tilde{k}_x (\diag[1, -1, -1, 1] \otimes \hat{\sigma}_z \ii\hat{\sigma}_y) \notag \\
 &\qquad + \sin\tilde{k}_y (\diag[1, -1, 1, -1] \otimes \hat{\sigma}_z \ii\hat{\sigma}_y) \bigr],
\end{align}
for the $A_1$ case, and
\begin{align}
 \hat{\Delta}_{\bm{Q}}^{\text{(init)}}(k) &\sim
 \sin\tilde{k}_x \sin\tilde{k}_y (\hat{\bm{1}}_4 \otimes \ii\hat{\sigma}_y) \notag \\
 &\quad + \delta \bigl[ \sin\tilde{k}_x (\diag[1, -1, 1, -1] \otimes \hat{\sigma}_z \ii\hat{\sigma}_y) \notag \\
 &\qquad + \sin\tilde{k}_y (\diag[1, -1, -1, 1] \otimes \hat{\sigma}_z \ii\hat{\sigma}_y) \bigr],
\end{align}
for the $A_2$ case, where the dominant $d$-wave and subdominant $p$-wave order parameter with $\delta = 0.1$ are taken into account.
Note that, due to the sublattice degree of freedom, the subdominant spin-triplet component can be admixed in the even-parity order parameter; in particular, we confirmed that the $p$-wave component has a comparatively large contribution when the spin splitting is present.
We here take $\tilde{\bm{k}} := \bm{k} - \bm{Q}/2$, since the order parameter obeys the same relation as Eq.~\eqref{eq:Green_func_anomalous_relation} for the anomalous Green's function, according to Eq.~\eqref{eq:Green_func_anomalous_2}.
In addition, we assume the even-frequency order parameter in all calculations.

By finding the optimal COM momentum(s) $\bm{Q}_{\text{opt}}$ such that the Eliashberg eigenvalue $\lambda_{\bm{Q}}$ reaches a maximum in the momentum space, we can draw the $h$--$n$ phase diagram as in Fig.~\ref{fig:phase_diagram_Hubbard}(a).
First, the $A_2$ order (circles) is more stable than the $A_1$ order (squares) in the wide range of parameters.
Second, the eigenvalue takes the peak at finite $\bm{Q}$s in the hole-doped AFM regime, which indicates that the FFLO superconductivity is stabilized when the spin splitting is large.
In Figs.~\ref{fig:phase_diagram_Hubbard}(b) and \ref{fig:phase_diagram_Hubbard}(c), for example, the momentum dependencies of $\lambda_{\bm{Q}}$ for $(n, h) = (1.3, 0)$ and $(1.3, 0.2)$ are exhibited, respectively.
The eigenvalue has a single peak at $\bm{Q} = \bm{0}$ in the paramagnetic state [Fig.~\ref{fig:phase_diagram_Hubbard}(b)], while the maximum is located at finite $\bm{Q}_{\text{opt}} \parallel \hat{x}$ in the AFM state [Fig.~\ref{fig:phase_diagram_Hubbard}(c)].
We also confirm that $\lambda_{\bm{Q}} = \lambda_{-\bm{Q}}$ due to the existence of the $C_{2z}$ symmetry.

Although the parameter region that we find the FFLO state is relatively limited compared to the model in Sec.~\ref{sec:effective_model}, we believe that the results showing its stability using a realistic model and reliable many-body methodologies is encouraging.
In recent numerical studies, the inclusion of off-site Coulomb repulsion terms, on top of the Hubbard model we considered, alters the competition of different symmetries and plays important roles in the SC instability, near $n = 1.5$ at $h = 0$~\cite{HiroshiWatanabe2017, HiroshiWatanabe2019}.
We expect modifications by the off-site terms for the finite AFM field $h$ as well, whose investigation is left for future studies.

\section{Discussion}
\label{sec:discussion}
We briefly discuss promising experimental setups to observe the AFM-induced FFLO superconductivity.
First, the coexistence of the AFM and SC orders was reported in previous experiments, where the phase separation appears in bulk $\kappa$-(ET)$_2$Cu[N(CN)$_2$]Br~\cite{Miyagawa2002, Sasaki2005, Fournier2007}.
In particular, the deuterated sample is located in the vicinity of the Mott transition and is a candidate system for the realization of the FFLO state in the anisotropic AFM spin splitting.
Another proposal is making use of recently developed experimental techniques to control phases by carrier doping and strain on a transistor device using a thin single crystal of $\kappa$-(ET)$_2X$~\cite{Suda2015, Kawasugi2019}.
If the AFM phase and the SC phase appear in adjacent regions of the sample, the coexistence can be realized via the proximity effect.

Furthermore, our results show the double-$\bm{Q}$ structure because of the twofold rotation (or inversion) symmetry.
Therefore, two possible forms of the order parameter are considered for the SC state below $T_c$: the single-$\bm{Q}$ (FF) state with $\Delta(\bm{r}) \sim \exp(\ii \bm{Q}_{\text{opt}} \cdot \bm{r})$ and the multi-$\bm{Q}$ (LO) state with $\Delta(\bm{r}) \sim \cos(\bm{Q}_{\text{opt}} \cdot \bm{r})$ ($\bm{r}$: real-space coordinate).
In the case of the Zeeman-field-induced FFLO state, clean isotropic $s$-wave superconductors prefer the multi-$\bm{Q}$ phase rather than the single-$\bm{Q}$ phase~\cite{Shimahara1998, Matsuda2007_review}, whereas both FF and LO states appear in anisotropic superconductors with nonmagnetic impurities~\cite{Agterberg2001}.
Since the impurity effect is generally considered to be small in organic conductors, we speculate that the multi-$\bm{Q}$ LO state is more stable in the coexisting state of the AFM and SC orders.
If the LO state is realized, the modulation of the SC order may be observable by scanning tunneling spectroscopy.
To actually determine which state is stabilized in the anisotropic splitting, solving the Bogoliubov--de Gennes equation or constructing the Ginzburg--Landau theory is needed, which is beyond the scope of the paper.

We here remark on the critical molecular field for the transition to the FFLO state.
In the case of the FFLO superconductivity stabilized by an external magnetic field, the transition occurs when the Zeeman field becomes comparable to the magnitude of the SC order parameter at zero temperature $\Delta(T = 0)$, which was shown by a mean-field analysis~\cite{Shimahara1994}.
A previous study on a single-band model introducing the anisotropic spin splitting demonstrated that the transition from a uniform $d$-wave SC state to a pair-density-wave (i.e., FFLO) state occurs also when the energy scale of the spin splitting approximately corresponds to $\Delta(0)$~\cite{Soto-Garrido2014}.
Therefore, one may expect that the critical AFM molecular field for the emergence of the FFLO state is given by $|h| \sim \Delta(0)$ in our case as well.
However, such a clear criterion is not easily seen here, because of the following three reasons.
First, the magnitude of the anisotropic spin splitting is not a simple function of $h$ since it sensitively depends on the carrier density, the real-space anisotropy of hopping integrals, and the resulting shape of Fermi surfaces.
Second, interband interactions as well as intraband ones alter the critical field, as indicated by the comparison between the effective model (Fig.~\ref{fig:phase_diagram_effective}) and the Hubbard model (Fig.~\ref{fig:phase_diagram_Hubbard}).
Third, the stability of the FFLO state is strongly related to the symmetry of the SC order parameter, as shown by the difference between Figs.~\ref{fig:phase_diagram_effective}(a) and \ref{fig:phase_diagram_effective}(b) in the effective model analysis.
These reasons are attributed to the multiband and anisotropic properties of our model, both of which are inherent in realistic altermagnetic materials.

Our theoretical proposal of the AFM-induced FFLO superconductivity is applicable not only to $\kappa$-type organic conductors but also to many other materials hosting the unusual $\bm{q} = 0$ AFM (or magnetic octupole/altermagnetic) order with anisotropic spin splitting, which have recently attracted much interest~\cite{Suzuki2017, Suzuki2019, Hayami2018, HikaruWatanabe2018, Smejkal2022_Sep, Smejkal2022_Dec}.
Indeed, a recent theoretical study~\cite{Zhang2023_arXiv} has proposed that Cooper pairs in altermagnets acquire finite COM momentum through the analysis of a Cooper-pair propagator~\cite{Lee1972} in a continuum model.
We should note that, unlike the case of the Zeeman field where highly two-dimensional compounds are required for the realization of the FFLO state to avoid orbital breaking, the mechanism here is free from such an effect;
therefore, three-dimensional compounds can also be candidates, expanding the scope of target materials.

\section{Summary}
\label{sec:summary}
In this paper, we microscopically investigated the property of superconductivity coexisting with the AFM order in (carrier-doped) $\kappa$-(ET)$_2X$.
By considering the effective theory with intraband attractive interactions, we found that FFLO superconductivity with finite COM momentum can be stabilized by the anisotropic spin splitting in the AFM phase.
Then we considered the on-site repulsive Hubbard model as a more realistic situation, and analyzed the linearized Eliashberg equation based on the FLEX approximation.
As a result, the FFLO state is realized in the hole-doped AFM state with comparatively large spin splitting.
In the whole calculations, we adopt the anisotropic order parameter with extended $s + d_{x^2-y^2}$-wave ($A_1$) or $d_{xy}$-wave ($A_2$) symmetry, the latter of which has nodes coinciding with the zeros of spin splitting and is more likely to induce the FFLO state.
This indicates that the anisotropic superconductivity is important for the appearance of the finite-momentum state in the anisotropic spin splitting~\cite{Soto-Garrido2014, Note5}.
\footnotetext{D.~F.~Agterberg (private communication).}

This finding paves a way to search for the FFLO state for the following reasons.
As we mentioned in Introduction, the occurrence of the FFLO superconductivity was originally suggested in high magnetic fields with the strong Zeeman effect, whereas a concomitant orbital pair breaking effect with vortex states obstructs the observation of the FFLO state.
On the other hand, the AFM-induced FFLO state is realized without an external magnetic field, and thus \textit{free from vortices}.
Therefore, the AFM state with spin splitting can be a good platform to realize the exotic FFLO superconductivity.
Another distinction of the AFM state from the Zeeman field is that the magnitude of the spin splitting strongly depends on the carrier density, even when the energy scale of the AFM molecular is unchanged.
Therefore, the spatial modulation of the FFLO order may be controllable by carrier doping.
Finally, we believe that our theory provides a foundation for further investigations of exotic superconductivity in the spin-splitting AFM state.

%%%%%%%%%% Acknowledgments %%%%%%%%%%
\begin{acknowledgments}
 The authors are grateful to Kosuke Nogaki and Youichi Yanase for technical advice on numerical calculations, and Hiroshi Watanabe for fruitful discussions about $\kappa$-type organic superconductors.
 The authors also thank Daniel F. Agterberg, Akito Daido, and Taisei Kitamura for valuable comments on this paper.
 This paper was supported by JSPS KAKENHI Grants No.~JP23K03333, No.~JP23K13056, No.~JP19K03723, No.~JP23H01129, No.~JP23H04047, and No.~JP20H04463, and the GIMRT Program of the Institute for Materials Research, Tohoku University, Grants No.~202112-RDKGE-0019 and No.~202212-RDKGE-0062.
\end{acknowledgments}

\appendix
\section{Symmetry of susceptibility}
\label{app:susceptibility_sym}
We show that the phase factor $\ee^{-\ii \bm{q} \cdot (\bm{r}_{l_1} - \bm{r}_{l_3})}$ in the multipole susceptibility [Eq.~\eqref{eq:multipole_susceptibility}] is necessary to satisfy the crystal symmetry of the system.
Conversely, the susceptibility breaks some (nonsymmorphic) symmetries if the phase factor is not taken into account, which can be understood by considering two distinct definitions of Fourier transformation.
Note that the discussions in this appendix have overlaps with Ref.~\cite{Nourafkan2017}.

\subsection{Preliminary: Fourier transformation}
We here introduce two different definitions of the Fourier transformation between the real and momentum spaces.
First, let $H_0$ be a tight-binding Hamiltonian with spin and sublattice degrees of freedom~\footnote{For simplicity, we neglect an orbital degree of freedom. However, the discussions in Appendix~\ref{app:susceptibility_sym} can be generalized even when the orbitals are taken into account.},
\begin{equation}
 H_0 = \sum_{\bm{R}, \bm{R}'} \sum_{l, l'} \sum_{s, s'} c_{\bm{R}, ls}^\dagger H_0(\bm{R} - \bm{R}')_{ls, l's'} c_{\bm{R}', l's'},
\end{equation}
where $c_{\bm{R}, ls}$ is the annihilation operator of an electron with spin $s$ at sublattice $l$ in unit cell $\bm{R}$.
Note that the electron is localized at the site (molecule) on $\bm{R} + \bm{r}_l$ in the real space.
Then we consider the Fourier transformation \textit{with} the internal coordinate $\bm{r}_l$ (Convention 1) defined by
\begin{equation}
 \tilde{c}_{\bm{k}, ls} = \sum_{\bm{R}} \ee^{-\ii\bm{k}\cdot(\bm{R}+\bm{r}_l)} c_{\bm{R}, ls}, \quad
 \tilde{c}_{\bm{R}, ls} = \frac{1}{N} \sum_{\bm{k}} \ee^{\ii\bm{k}\cdot(\bm{R}+\bm{r}_l)} c_{\bm{k}, ls},
\end{equation}
where the tilde on a character means being associated with Convention 1, which is the case for other operators in the following.
In this Convention, the Hamiltonian is given by
\begin{equation}
 H_0 = \frac{1}{N} \sum_{\bm{k}} \sum_{l, l'} \sum_{s, s'} \tilde{c}_{\bm{k}, ls}^\dagger \tilde{H}_0(\bm{k})_{ls, l's'} \tilde{c}_{\bm{k}, l's'},
\end{equation}
where the momentum-space Hamiltonian
\begin{equation}
 \tilde{H}_0(\bm{k})_{ls, l's'} = \sum_{\bm{R}} \ee^{-\ii\bm{k}\cdot(\bm{R}+\bm{r}_l-\bm{r}_{l'})} H_0(\bm{R})_{ls, l's'}
\end{equation}
is not invariant under the transformation $\bm{k} \mapsto \bm{k} + \bm{G}$, with $\bm{G}$ being a reciprocal lattice vector.
In other words, the Hamiltonian matrix $\hat{\tilde{H}}_0(\bm{k})$, in general, breaks the periodicity of the Brillouin zone.

Another definition of the Fourier transformation \textit{without} the internal coordinate (Convention 2) is given by
\begin{equation}
 c_{\bm{k}, ls} = \sum_{\bm{R}} \ee^{-\ii\bm{k}\cdot\bm{R}} c_{\bm{R}, ls}, \quad
 c_{\bm{R}, ls} = \frac{1}{N} \sum_{\bm{k}} \ee^{\ii\bm{k}\cdot\bm{R}} c_{\bm{k}, ls},
\end{equation}
(tilde is not used).
The Hamiltonian in Convention 2 is
\begin{gather}
 H_0 = \frac{1}{N} \sum_{\bm{k}} \sum_{l, l'} \sum_{s, s'} c_{\bm{k}, ls}^\dagger H_0(\bm{k})_{ls, l's'} c_{\bm{k}, l's'}, \\
 H_0(\bm{k})_{ls, l's'} = \sum_{\bm{R}} \ee^{-\ii\bm{k}\cdot\bm{R}} H_0(\bm{R})_{ls, l's'}.
\end{gather}
Therefore, the Hamiltonian matrix $\hat{H}_0(\bm{k})$ is invariant under $\bm{k} \mapsto \bm{k} + \bm{G}$, and thus satisfies the periodicity of the Brillouin zone.
This property is compatible with the calculation of a topological invariant and the fast Fourier transformation (FFT) technique.
In this paper, we basically adopt Convention 2 for the formulation of the Hamiltonian [Eq.~\eqref{eq:Hamiltonian_momentum}], since we need to use the FFT in numerical calculations of Green's functions in Sec.~\ref{sec:Eliashberg_Hubbard}.

Now we consider the transformation property under symmetry operations.
Let $g = \{p_g | \bm{a}_g\} \in G$ be a symmetry operator in space group $G$ of the system, where $\{p_g | \bm{a}_g\}$ is the Seitz notation with $p_g$ and $\bm{a}_g$ being the point group operation and translation, respectively.
The electron annihilation operator in the real space transforms under $g$ as
\begin{align}
 g c_{\bm{R}, ls} g^{-1}
 &= \sum_{s'} c_{\bm{R}', gl, s'} D^{\text{(spin)}}(g)_{s', s} \notag \\
 & (\bm{R}' := p_g \bm{R} + p_g \bm{r}_l + \bm{a}_g - \bm{r}_{gl}),
\end{align}
where $\hat{D}^{\text{(spin)}}(g)$ is a unitary representation matrix of $g$ with respect to the spin degree of freedom.
Using the relation, we can derive the transformation property of the creation operator in the momentum space:
\begin{align}
 g \tilde{c}_{\bm{k}, l s} g^{-1}
 &= \ee^{\ii p_g\bm{k} \cdot \bm{a}_g} \sum_{s'} \tilde{c}_{p_g\bm{k}, gl, s'} D^{\text{(spin)}}(g)_{s', s}
 \label{eq:transform_ck_1} \\
 \intertext{in Convention 1, and}
 g c_{\bm{k}, l s} g^{-1}
 &= \ee^{\ii p_g\bm{k} \cdot (\bm{a}_g + p_g \bm{r}_l - \bm{r}_{gl})} \sum_{s'} c_{p_g\bm{k}, gl, s'} D^{\text{(spin)}}(g)_{s', s}
 \label{eq:transform_ck_2}
\end{align}
in Convention 2.
Depending on the Conventions, the phase factors in the transformation are different.
Also, the unitary matrix representing the symmetry operation $g$ is defined by
\begin{align}
 \tilde{U}_g(\bm{k})_{ls, l's'} &:= \ee^{\ii p_g\bm{k} \cdot \bm{a}_g} \delta_{l, gl'} D^{\text{(spin)}}(g)_{s, s'} \\
 \intertext{in Convention 1, and}
 U_g(\bm{k})_{ls, l's'} &:= \ee^{\ii p_g\bm{k} \cdot (\bm{a}_g + p_g \bm{r}_l - \bm{r}_{gl})} \delta_{l, gl'} D^{\text{(spin)}}(g)_{s, s'}
 \label{eq:symmetry_unitary_conv2}
\end{align}
in Convention 2.
Equation~\eqref{eq:symmetry_unitary_conv2} corresponds to the unitary matrices in Eqs.~\eqref{eq:symmetry_unitary} of the main text.

\subsection{Susceptibility}
Based on the above preliminaries, we demonstrate that the dynamical susceptibility breaks some symmetries in Convention 2, whereas that in Convention 1 preserves all the symmetries.
Here we choose the longitudinal FM (LFM) spin susceptibility for the model of $\kappa$-(ET)$_2X$ that we treated in the main text with $G = p2gg$, as an example; the generalization to other susceptibilities and/or symmetries straightforward.
For this purpose, the spin operator on the sublattice $l$ in the unit cell $\bm{R}$ is defined by
\begin{equation}
 s^i_l(\bm{R}) := \sum_{s_1, s_2} c_{\bm{R}, l s_1}^\dagger \sigma^i_{s_1, s_2} c_{\bm{R}, l s_2}
 \quad (i = x, y, z).
\end{equation}
In the momentum space, the spin operator is transformed as
\begin{align}
 \tilde{s}^i_l(\bm{q})
 &= \sum_{\bm{R}} \ee^{-\ii \bm{q} \cdot (\bm{R} + \bm{r}_l)} s^i_l(\bm{R})
 = \frac{1}{N} \sum_{\bm{k}} \sum_{s_1, s_2} \tilde{c}_{\bm{k}, l s_1}^\dagger \sigma^i_{s_1, s_2} \tilde{c}_{\bm{k} + \bm{q}, l s_2} \\
 \intertext{in Convention 1, and}
 s^i_l(\bm{q})
 &= \sum_{\bm{R}} \ee^{-\ii \bm{q} \cdot \bm{R}} s^i_l(\bm{R})
 = \frac{1}{N} \sum_{\bm{k}} \sum_{s_1, s_2} c_{\bm{k}, l s_1}^\dagger \sigma^i_{s_1, s_2} c_{\bm{k} + \bm{q}, l s_2}
\end{align}
in Convention 2.
Using Eqs.~\eqref{eq:transform_ck_1} and \eqref{eq:transform_ck_2}, we next derive the transformation property of the longitudinal ($z$ direction) spin under $g \in G$.
After some algebra, it is given by
\begin{align}
 g \tilde{s}^z_l(\bm{q}) g^{-1}
 &= \ee^{\ii p_g\bm{q} \cdot \bm{a}_g} \tilde{s}^z_{gl}(p_g\bm{q}) D_{B_2}(p_g)
 \label{eq:transform_sz_1} \\
 \intertext{in Convention 1 and}
 g s^z_l(\bm{q}) g^{-1}
 &= \ee^{\ii p_g\bm{q} \cdot (\bm{a}_g + p_g \bm{r}_l - \bm{r}_{gl})} s^z_{gl}(p_g\bm{q}) D_{B_2}(p_g)
 \label{eq:transform_sz_2}
\end{align}
in Convention 2, where $D_{B_2}(p_g)$ represents the character of the $B_2$ irrep of the point group $C_{2v}$ (see Table~\ref{tab:character_C2v}), since the spin matrix transforms as $\hat{\sigma}_z \xmapsto{E, C_{2z}} \hat{\sigma}_z$ and $\hat{\sigma}_z \xmapsto{G_x, G_y} -\hat{\sigma}_z$.

\begin{figure}
 \includegraphics[width=\linewidth, pagebox=artbox]{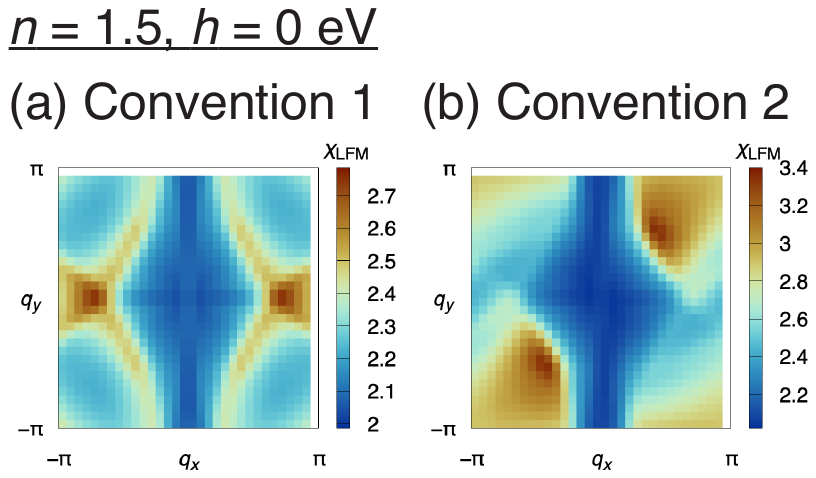}
 \caption{The LFM spin susceptibility defined for $n = 1.5$ and $h = 0$ in (a) Convention 1 [Eq.~\eqref{eq:susceptibility_1}] and (b) Convention 2 [Eq.~\eqref{eq:susceptibility_2}]. In (b), the susceptibility breaks the glide symmetries $G_x$ and $G_y$.}
 \label{fig:susspin_sym_asym}
\end{figure}

Using these spin operators, the LFM spin susceptibility is defined by
\begin{align}
 \tilde{\chi}_{\text{LFM}}(\bm{q}, \ii\omega_n)
 &:= \sum_{l, l'} \int_{0}^{\beta} \mathrm{d}\tau \, \ee^{i\omega_n\tau} \Braket{T_\tau \tilde{s}^z_l(\bm{q}, \tau) \tilde{s}^z_{l'}(-\bm{q}, 0)}
 \label{eq:susceptibility_1} \\
 \intertext{in Convention 1, and}
 \chi_{\text{LFM}}(\bm{q}, \ii\omega_n)
 &:= \sum_{l, l'} \int_{0}^{\beta} \mathrm{d}\tau \, \ee^{i\omega_n\tau} \Braket{T_\tau s^z_l(\bm{q}, \tau) s^z_{l'}(-\bm{q}, 0)}
 \label{eq:susceptibility_2}
\end{align}
in Convention 2.
Therefore, we just have to investigate the transformation property of $\sum_{l, l'} \tilde{s}^z_l(\bm{q}) \tilde{s}^z_{l'}(-\bm{q})$ [$\sum_{l, l'} s^z_l(\bm{q}) s^z_{l'}(-\bm{q})$] under $g$ in Convention 1 (2), using the relations Eqs.~\eqref{eq:transform_sz_1} and \eqref{eq:transform_sz_2}.
Then we obtain the following equation in Convention 1:
\begin{align}
 \tilde{\chi}_{\text{LFM}}(\bm{q}, \ii\omega_n)
 &= \sum_{l, l'} \int_{0}^{\beta} \mathrm{d}\tau \, \ee^{i\omega_n\tau} \Braket{T_\tau \tilde{s}^z_l(p_g\bm{q}, \tau) \tilde{s}^z_{l'}(-p_g\bm{q}, 0)} \notag \\
 &= \tilde{\chi}_{\text{LFM}}(p_g\bm{q}, \ii\omega_n), \ \forall g \in p2gg,
 \label{eq:transform_chi_1}
\end{align}
which indicates that the susceptibility is invariant under any operation in point group $C_{2v}$.
In Convention 2, on the other hand, the similar calculations result in
\begin{align}
 \chi_{\text{LFM}}(\bm{q}, \ii\omega_n)
 &= \sum_{l, l'} \ee^{\ii p_g\bm{q} \cdot (p_g \bm{r}_l - \bm{r}_{gl} - p_g \bm{r}_{l'} + \bm{r}_{gl'})} \notag \\
 &\quad \times \int_{0}^{\beta} \mathrm{d}\tau \, \ee^{i\omega_n\tau} \Braket{T_\tau s^z_l(p_g\bm{q}, \tau) s^z_{l'}(-p_g\bm{q}, 0)},
\end{align}
where the phase factor $\ee^{\ii p_g\bm{q} \cdot (p_g \bm{r}_l - \bm{r}_{gl} - p_g \bm{r}_{l'} + \bm{r}_{gl'})}$ has a nontrivial contribution for the nonsymmorphic symmetries $g = G_x$ and $G_y$, while it is identity for $g = E$ and $C_{2v}$.
Therefore, the susceptibility in Convention 2 breaks $G_x$ and $G_y$.
Indeed, we can see the symmetry properties of the LFM spin susceptibility displayed in Fig.~\ref{fig:susspin_sym_asym}.
In Convention 1, all the $C_{2v}$ symmetry is conserved [Fig.~\ref{fig:susspin_sym_asym}(a)], whereas the two glide symmetries are broken in Convention 2 [Fig.~\ref{fig:susspin_sym_asym}(b)].

Finally, we explain how to recover the broken symmetry of the susceptibility calculated in Convention 2.
By using the relation
\begin{equation}
 \tilde{s}^i_l(\bm{q}) \tilde{s}^j_{l'}(-\bm{q})
 = \ee^{-\ii\bm{q} \cdot (\bm{r}_l - \bm{r}_{l'})} s^i_l(\bm{q}) s^j_{l'}(-\bm{q}),
\end{equation}
the spin susceptibility in Convention 1 [Eq.~\eqref{eq:susceptibility_1}] is rewritten as
\begin{align}
 \tilde{\chi}_{\text{LFM}}(\bm{q}, \ii\omega_n)
 &= \sum_{l, l'} \ee^{-\ii\bm{q} \cdot (\bm{r}_l - \bm{r}_{l'})} \notag \\
 &\quad \times \int_{0}^{\beta} \mathrm{d}\tau \, \ee^{i\omega_n\tau} \Braket{T_\tau s^z_l(\bm{q}, \tau) s^z_{l'}(-\bm{q}, 0)}.
 \label{eq:susceptibility_prescription}
\end{align}
Note that the spin operators in the integrand have no tilde symbol.
Therefore, the broken (nonsymmorphic) symmetry can be recovered by multiplying the phase factor $\ee^{-\ii\bm{q} \cdot (\bm{r}_l - \bm{r}_{l'})}$ to the integral calculated in Convention 2.
That is the reason we include the phase factor in Eq.~\eqref{eq:multipole_susceptibility}.
In all calculations of the susceptibility except Fig.~\ref{fig:susspin_sym_asym}(b), we use the internal coordinates $\bm{r}_A = (0, 0)$, $\bm{r}_B = (1/2, 1/2)$, $\bm{r}_C = (1/2, 0)$, and $\bm{r}_D = (0, 1/2)$, where the four sublattices (molecules) are arranged like the Shastry--Sutherland lattice~\cite{Seo2021}.

\section{Results in undoped regime}
\label{app:undoped}
\begin{figure}
 \includegraphics[width=\linewidth, pagebox=artbox]{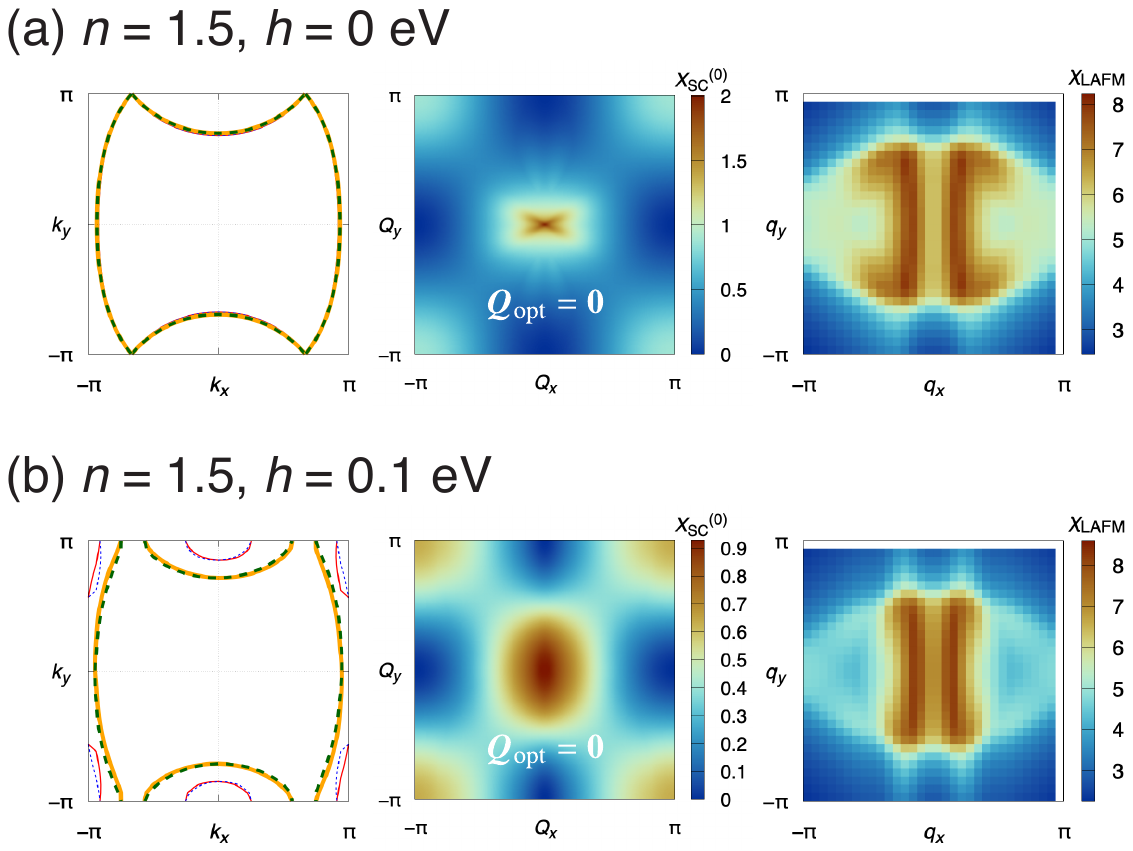}
 \caption{Results in the undoped case ($n = 1.5$) for (a) $h = 0$ eV and (b) $h = 0.1$ eV. The temperature $T$ is set to 1 meV. Left panels: The renormalized Fermi surfaces (orange solid and green dashed lines) and the noninteracting Fermi surfaces (red solid and blue dashed lines). Middle and right panels: $X_{\text{SC}}^{(0)}(\bm{Q})$ and the LAFM spin susceptibility.}
 \label{fig:fs_susSC_susspin_undoped}
\end{figure}

We here show the analyses in the undoped case ($n = 1.5$), while the electron- and hole-doped regimes are discussed in the main text.
Figures~\ref{fig:fs_susSC_susspin_undoped}(a) and \ref{fig:fs_susSC_susspin_undoped}(b) represent numerical results in the paramagnetic ($h = 0$ eV) and AFM metal ($h = 0.1$ eV) phases, respectively.
First, the calculations of $X_{\text{SC}}^{(0)}(\bm{Q})$ within the effective intraband model (Sec.~\ref{sec:effective_model}) are given by the middle panels, in which the $d_{xy}$-wave ($A_2$) interaction is assumed.
The SC susceptibility reaches a maximum at $\bm{Q} = 0$ in both cases, which indicates the conventional BCS state is stable even when the AFM order coexists.
Second, in the right panels of Fig.~\ref{fig:fs_susSC_susspin_undoped}, the LAFM spin susceptibility in the Hubbard model (Sec.~\ref{sec:Eliashberg_Hubbard}) is plotted.
We also confirmed that the structure of the FM spin susceptibility for $h = 0$ [Fig.~\ref{fig:susspin_sym_asym}(a)] is consistent with previous theoretical results based on the RPA~\cite{Guterding2016} and the FLEX approximation~\cite{Kuroki2002}.
The correlation effect largely appears for $h = 0.1$ eV as a renormalization of the Fermi surfaces, while such an effect is vanishingly small for $h = 0$ (see the left panels).

\section{LAFM and TAFM spin fluctuations}
\label{app:LAFM_TAFM}
\begin{figure*}
 \includegraphics[width=\linewidth, pagebox=artbox]{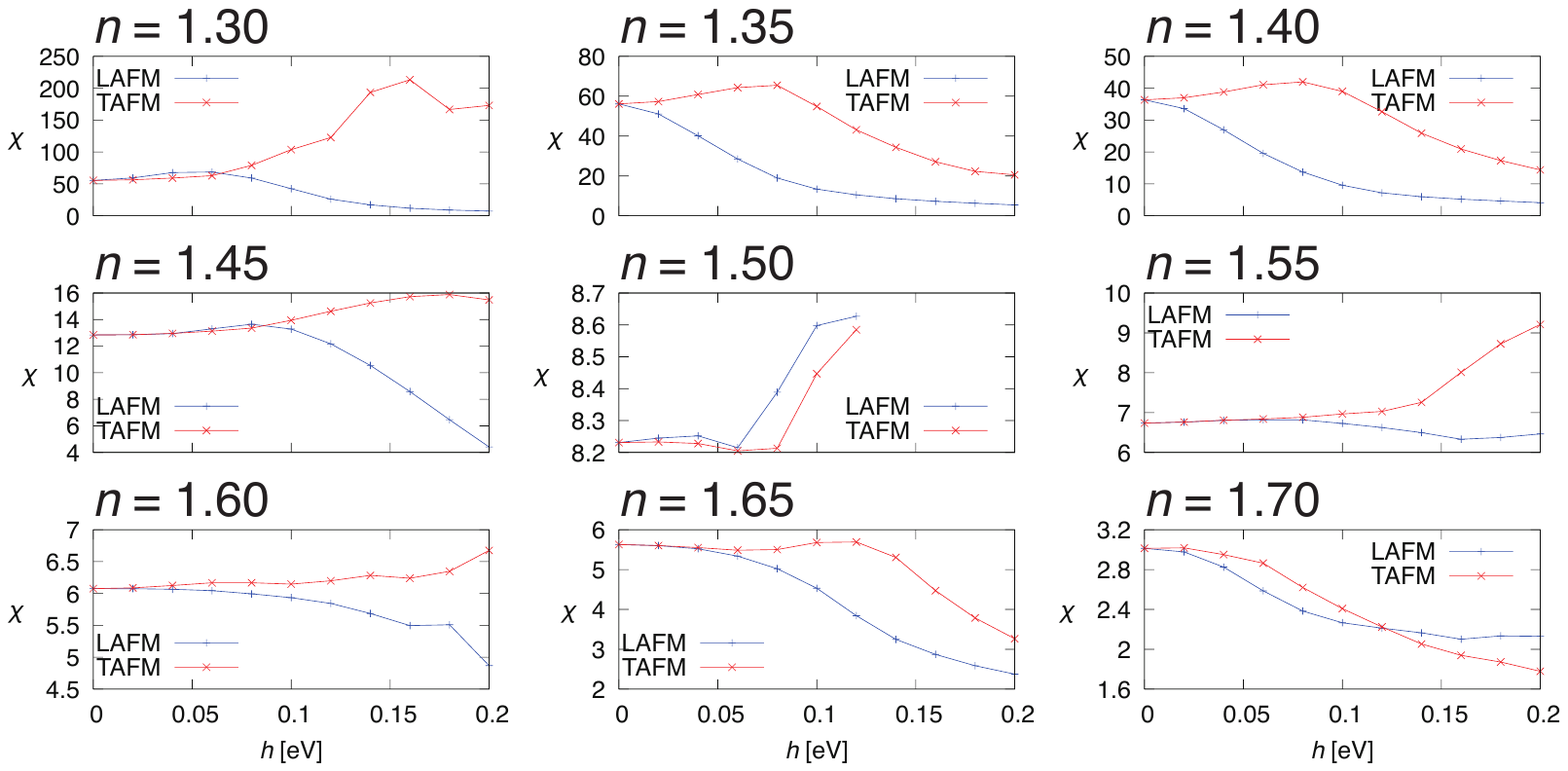}
 \caption{The molecular field $h$ dependence of the AFM fluctuations for each filling. The maxima of the LAFM and TAFM spin susceptibilities are shown. Both are degenerate for $h = 0$ due to the absence of the spin--orbit coupling.}
 \label{fig:sus_graph}
\end{figure*}

We show the results of the LAFM and TAFM spin susceptibilities in Fig.~\ref{fig:sus_graph}, in the Hubbard model (Sec.~\ref{sec:Eliashberg_Hubbard}) for several parameter sets.
In particular, the TAFM fluctuations for the large $h$ in the hole-doped regime are strongly enhanced, where the momentum dependence is exhibited in Fig.~\ref{fig:fs_susspin}(d).

% \bibliography{paper}

% 

\end{document}